\documentclass[12pt,amsmath,amssymb]{iopart}
\usepackage[utf8]{inputenc}
\usepackage{graphicx}

\newcommand{\christoffel}[2]{\Gamma^{#1}_{\hphantom{#1}#2}}

\usepackage{iopams}
\usepackage[%
breaklinks=true,
colorlinks=true,
linkcolor=magenta,
urlcolor=blue,
citecolor=blue
]{hyperref}

\begin{document}
\title[Equivalence between Scalar-Tensor theories and $f(R)$-gravity]{Equivalence between Scalar-Tensor theories and $f(R)$-gravity: From the action to Cosmological Perturbations}
\author{Joel Velásquez$^1$, Leonardo Castañeda$^1$}
\address{$^1$Grupo de Gravitación y Cosmología, Observatorio Astronómico Nacional\\Universidad Nacional de Colombia,  cra 45 \#26-85,Ed.Uriel Gutiérrez,\\ Bogotá D.C., Colombia}
\ead{lcastanedac@unal.edu.co, jjvelasquezc@unal.edu.co}

\begin{abstract}
	In this paper we calculate the field equations for Scalar-Tensor from a variational principle, taking into account the Gibbons-York-Hawking type boundary term. We do the same for the theories $f(R)$, following~\cite{Alejo}. Then, we review the equivalences between both theories in the metric formalism. Thus, starting from the perturbations for Scalar-Tensor theories, we find the perturbations for $f(R)$ gravity under the equivalences. Working with two specific models of $f(R)$, we explore the equivalences between the theories under conformal-Newtonian gauge. Further, we show the perturbations for both theories under the sub-horizon approach.
\end{abstract}


\section{\label{sec::Intro}Introduction}
% Put \label in argument of \section for cross-referencing
%\section{\label{}}
Recent observations of the CMB show that the universe is in accelerated expansion~\cite{Riess,Perlmutter}. The broadly used model is the $\Lambda$-Cold-Dark-Matter ($\Lambda$CDM). However, this model introduces an exotic term of energy, called Dark Energy (DE), associated to the cosmological constant term $\Lambda$. Assuming that the theory of general relativity (GR) is not entirely correct at cosmological scales, it is possible that a cosmological constant term is not necessary to explain the accelerated expansion of the universe. The alternative theories to the Einstein's proposal are known as modified gravity theories (MG). One set of these theories is known as Scalar-Tensor gravity theories (ST) ~\cite{Maheda_book, Capozziello_book, Venturi}, where the gravitational action in these theories, in addition to the metric, to contain a scalar field which intervenes in the generation of the space-time curvature, associated to the metric. This scalar field is not directly coupled to the matter and, therefore, the matter responds only to the metric. It should be noted that the Brans-Dicke theory (BD), ~\cite{Brans1} proposed by C.H. Brans y R.H. Dicke in 1961, is a particular case of theories ST, where the parameter $\omega(\phi)$ is independent of the scalar field.\\
Another type of generalization to GR are the theories of gravity $f(R)$ ~\cite{Sotiriou,DeFelice,Nojiri_2}, where the lagrangian of Einstein-Hilbert is generalized, replacing the scalar curvature $R$ by a more general function of it, $f(R)$. The gravitational field in this theory is represented by the metric like GR does.\\
The equivalence between theories ST and $f(R)$ has been studied e.g., in ~\cite{Belenchia,Capone,Ntahompagaze,Sotiriou-1,Odinstov,Nojiri}. It is, starting from the ST action, without the kinetic term of the scalar field, we arrive at the action of the gravity theories $f(R)$. In this paper, in addition to the above, we show these equivalences for the field equations, the Friedmann equations of the homogeneous and isotropic universe and the Friedmann's perturbations in any gauge. Further, we show two specific examples of theories $f(R)$ under the conformal-Newtonian gauge. \\
The paper is organized as following: in the section \ref{sec::Campo} we get the field equations for RG, ST and $f(R)$ theories starting from the variational principle, taking into account the Gibbons-York-Hawking (GYH) boundary term type, for every of the above theories. It is found that the consideration to obtain the field equations for ST, under the equivalence of the theories, to coincide to the $f(R)$ condition. In the section \ref{sec::equivalencias} the equivalence between ST and $f(R)$  for the actions and the field equations of the theories is shown. In the section \ref{sec::perturbaciones} the Friedmann equations for the background universe (homogeneous and isotropic) are calculated. Besides, we calculate the perturbed Friedmann equations, for ST, and the ones for $f(R)$, using the equivalence between the theories. Then, we show how to construct the potential for the Hu-Sawicki and Starobisnky $f(R)$ models, in order to calculate the Friedmann equations for the background and perturbed universe in these models for the two formalisms under the confomal-Newtonian gauge. Inmediately, we perform the sub-horizon approach to the perturbations, for both theories, and we show that they can not be calculated using the equivalences, due to the parameter $\omega=cte$ for ST. Finally, in the section \ref{sec::conclusiones} we show the conclusions. In the appendix \ref{sec::xPand} we show how the perturbations were calculated under the package \textit{xPand} from software \textbf{Mathematica}.\\
Throughout the review, we adopt natural units $8\pi G=c=1$, here $G$ is Newton’s gravitational constant and $c$ is speed of light. Have a metric signature $(-+++)$. Small latin indices $a$, $b$, $\ldots$ assume the values 0 to 3, while greek indices $\alpha$, $\beta$, $\ldots$ assume the values 1,2,3.
%============================================================================================================================================================================================================================================
\section{\label{sec::Campo}Field equations and Variational principles}
This section shows how the field equations, through a variational principle for the theories GR, ST and $f(R)$ are found; taking into account in all of these theories the boundary term type GYH.
.
%=============================================================================================================================================================
\subsection{Field Equations in GR}
The Einstein field equations (EFEs) can be deduced through a variational principle. We give a detailed review following ~\cite{Wald_book,Poisson_book,Carroll_book}. The action for GR is
\begin{equation}\label{eq:accionRG}
	S^{(RG)}=\frac{1}{16\pi}\int_\mathcal{M}d^{4}x\sqrt{-g}R+S^{(m)}(g_{ab},\psi),
\end{equation}
where the first term is known as the Einstein-Hilbert action,  $d^{4}x\sqrt{-g}$ is the element of invariant volume and $R$ is the  Ricci scalar.\\
\noindent The second term is the matter action defined by
\begin{equation}
	\label{eq:accionM}
	S^{(m)}=\int_{\mathcal{M}}d^{4}x\sqrt{-g}\mathcal{L}^{(m)}(g_{ab},\psi),
\end{equation}
where $\psi$ denotes the matter fields.\\ 
The variation of the action (\ref{eq:accionRG}) with respect to $g^{ab}$ takes the form
\begin{equation}\label{eq:varaccionRG1}
	\delta S^{(RG)}=\frac{1}{16\pi}\int_\mathcal{M}d^{4}x\delta(\sqrt{-g} R)+\delta S^{(m)}.
\end{equation}
Given the variation of the Ricci scalar
\begin{equation}\label{eq:deltaR}
	\delta R=\delta g^{ab}R_{ab}+\nabla_{c}(g^{ab}\delta\christoffel{c}{ab})-\nabla_{b}(g^{ab}\delta\christoffel{c}{ac}),
\end{equation}
we get
\begin{equation}\label{eq:varaccionRG2}
	\fl\delta S^{(RG)}=\frac{1}{16\pi}\int_\mathcal{M}d^{4}x\sqrt{-g}\big(R_{ab}-\frac{1}{2}g_{ab}R\big)\delta g^{ab}
	+\frac{1}{16\pi}\int_\mathcal{M}d^{4}x\sqrt{-g}\nabla_{d}V^{d}+\delta S^{(m)},
\end{equation}
where
\begin{equation}\label{eq:VectorV}
	V^{d}=g^{ab}\delta\christoffel{d}{ab}-g^{ad}\delta\christoffel{c}{ac}.
\end{equation}
The second integral of the equation (\ref{eq:varaccionRG2}) is a divergence term. Thus, we can use the Gauss-Stokes theorem 
\begin{equation}\label{eq:gauss-stokes}
	\int_{\mathcal{M}}d^{4}x\sqrt{|g|}\nabla_{d}A^{d}=\oint_{\partial\mathcal{M}}d^{3}y\epsilon\sqrt{|h|}n_{d}A^{d},
\end{equation}
where $\partial\mathcal{M}$ its the boundary of a hypervolume on $\mathcal{M}$, $h$ is the determinant of the induced metric, $n_{d}$ is the unit normal vector to $\partial\mathcal{M}$, $\epsilon$ is $+1$ if $\partial\mathcal{M}$ is timelike and $-1$ if
$\partial\mathcal{M}$ is spacelike (it is assumed that $\partial\mathcal{M}$ is nowhere null). Coordinates $x^{a}$ are used for the finite region $\mathcal{M}$ and $y^{a}$ for the boundary $\partial\mathcal{M}$.\\
In the equation (\ref{eq:VectorV}) the variations of the Christoffel symbols are present. Calculating this variations in the boundary, we have
\begin{equation}\label{eq:varchristoffel2}
	\delta\christoffel{a}{bc}\Big\vert_{\partial\mathcal{M}}=\frac{1}{2}g^{ad}(\partial_{b}\delta g_{dc}+\partial_{c}\delta g_{bd}-\partial_{d}\delta g_{bc}),
\end{equation}
where it has been imposed that the variation of the metric tensor is null in the boundary, i.e.,
\begin{equation}\label{eq:condvarmetrica}
	\delta g_{ab}\Big\vert_{\partial\mathcal{M}}=0.
\end{equation}
Found the equation (\ref{eq:varchristoffel2}), the vector $V_{d}=g_{ed}V^{e}$ is calculated at the boundary
\begin{equation}
	V_{d}\Big\vert_{\partial\mathcal{M}}=g^{ab}(\partial_{b}\delta g_{da}-\partial_{d}\delta g_{ba}).
\end{equation}
Now we evaluate the term $n^{d}V_{d}\big\vert_{\partial\mathcal{M}}$, using for this
\begin{equation}\label{eq:metricafrontera}
	g^{ab}=h^{ab}+\epsilon n^{a}n^{b},
\end{equation}
then
\begin{equation}
	n^{d}V_{d}\big\vert_{\partial\mathcal{M}}=n^{d}h^{ab}(\partial_{b}\delta g_{da}-\partial_{d}\delta g_{ba}),
\end{equation}
where we use the antisymmetric part of $\epsilon n^{a}n^{b}$, with $\epsilon=n^{d}n_{d}=\pm1$. To the fact $\delta g_{ab}=0$ in the boundary we have $h^{ab}\partial_{b}\delta g_{da}=0$, we get
\begin{equation}
	n^{d}V_{d}\big\vert_{\partial\mathcal{M}}=-n^{d}h^{ab}\partial_{d}\delta g_{ba}.
\end{equation}
The variation of the action (\ref{eq:varaccionRG2}) takes the form
\begin{equation}\label{eq:varaccionRG3}
\fl\delta S^{(RG)}=\frac{1}{16\pi}\int_\mathcal{M}d^{4}x\sqrt{-g}\big(R_{ab}-\frac{1}{2}g_{ab}R\big)\delta g^{ab}
	-\frac{1}{16\pi}\oint_{\partial\mathcal{M}}d^{3}y\epsilon\sqrt{|h|}n^{d}h^{ab}\partial_{d}\delta g_{ba}+\delta S^{(m)}.
\end{equation}
The above equation shows that fixing $\delta g_{ab}=0$ on $\partial\mathcal{M}$ there is an additional boundary term. It could be argued that both the variation of the metric and its first derivative vanish in the boundary, i.e., $\delta g_{ab}=0$ and $\partial_{c}\delta g_{ab}=0$ in $\partial\mathcal{M}$. Although this last argument leads directly to Einstein field equations, it implies to fix two conditions in the boundary. To avoid this, a boundary term is introduced, the Gibbons-York-Hawking (GYH) boundary term, 
that allows to have a well defined variational problem only  fixing the variation of the metric in the boundary, $\delta g_{ab}\big\vert_{\partial\mathcal{M}}=0$ ~\cite{Gibbons,Hawking}. This term is
\begin{equation}\label{eq:GYH-RG}
	S^{(RG)}_{GYH}=\frac{1}{8\pi}\oint_{\partial\mathcal{M}}d^{3}y\epsilon\sqrt{|h|} K,
\end{equation}
where $K$ is the trace of extrinsic curvature. The variation of the GYH action is
\begin{equation}\label{eq:varGYH-RG}
	\delta S^{(RG)}_{GYH}=\frac{1}{8\pi}\oint_{\partial\mathcal{M}}d^{3}y\epsilon\sqrt{|h|} \delta K,
\end{equation}
where $\delta h^{ab}=0$ in the boundary $\partial\mathcal{M}$.\\
Using the definition of the extrinsic curvature ~\cite{Poisson_book}
\begin{equation}
	K_{ab}=h_{a}^{\ c}\nabla_{c}n_{b},
\end{equation}
the trace is given by
\begin{equation}
	K=\nabla_{a}n^{a}=g^{ab}\nabla_{b}n_{a}=h^{ab}(\partial_{b}n_{a}-\christoffel{c}{ba}n_{c}),
\end{equation}
where we have used the equation (\ref{eq:metricafrontera}). Taking into account (\ref{eq:varchristoffel2}), $\delta K$ is calculated on the boundary
\begin{equation}\label{eq:vartrazak}
	\delta K=-h^{ab}\delta\christoffel{c}{ba}n_{c}=\frac{1}{2}h^{ab}\partial_{d}\delta g_{ba}n^{d}.
\end{equation}
The variation (\ref{eq:varGYH-RG}) gives
\begin{equation}\label{eq:varGYH-RG2}
	\delta S^{(RG)}_{GYH}=\frac{1}{16\pi}\oint_{\partial\mathcal{M}}d^{3}y\epsilon\sqrt{|h|}h^{ab}\partial_{d}\delta g_{ba}n^{d}.
\end{equation}
This term to cancel with the second integral of (\ref{eq:varaccionRG3}) (the boundary term contribution). Hence we have
\begin{equation}\label{eq:varaccionRG4}
	\delta S^{(RG)}=\frac{1}{16\pi}\int_\mathcal{M}d^{4}x\sqrt{-g}\big(R_{ab}-\frac{1}{2}g_{ab}R\big)\delta g^{ab}+\delta S^{(m)}.
\end{equation}
The variation of the action (\ref{eq:accionM}) takes the form
\begin{equation}
	\fl\delta S^{(m)}=\int_{\mathcal{M}}d^{4}x\delta(\sqrt{-g}\mathcal{L}^{(m)})
	=\int_{\mathcal{M}}d^{4}x\sqrt{-g}\left(\frac{\partial\mathcal{L}^{(m)}}{\partial g^{ab}}-\frac{1}{2}\mathcal{L}^{(m)}g_{ab}\right)\delta g^{ab}.
\end{equation}
Defining the stress-energy tensor by
\begin{equation}
	T_{ab}\equiv-2\frac{\partial\mathcal{L}^{(m)}}{\partial g^{ab}}+\mathcal{L}^{(m)}g_{ab}=-\frac{2}{\sqrt{-g}}\frac{\delta S^{(m)}}{\delta g^{ab}},
\end{equation}
then
\begin{equation}\label{eq:varaccionM}
	\delta S^{(m)}=-\frac{1}{2}\int_{\mathcal{M}}d^{4}x\sqrt{-g}T_{ab}\delta g^{ab}.
\end{equation}
Imposing that the total variations to remain invariant with respect to $\delta g^{ab}$, i.e.,
\begin{equation}
	\frac{1}{\sqrt{-g}}\frac{\delta S^{(RG)}}{\delta g^{ab}}=0.
\end{equation}
Finally, we get 
\begin{equation}\label{eq:campoRG}
	R_{ab}-\frac{1}{2}Rg_{ab}=8\pi T_{ab},
\end{equation}
which are the Einstein field equations.
%=============================================================================================================================================================
\subsection{Field Equations in ST gravity}
Scalar-Tensor theories of gravity belong to the MG theories, where a function of scalar field $\phi$ is non-minimal coupling to the Ricci scalar $R$. The action in the so-called Jordan Frame is ~\cite{faraoni_book}
\begin{equation}
	\label{eq:accionST}
	S^{(ST)}=
	\int_{\mathcal{M}}d^{4}x\sqrt{-g}\left[\frac{f(\phi)}{2}
	R-\frac{\omega(\phi)}{2}g^{ab}\nabla_{a}\phi\nabla_{b}\phi-V(\phi)\right]\quad+S^{(m)},
\end{equation}
where $S^{(m)}$ is the action (\ref{eq:accionM}) describing ordinary matter (any form of matter different from the scalar field $\phi$), $\omega$ is a parameter that is a function of the scalar field $\phi$. Notice that the matter is not directly coupled to $\phi$, in the sense that the Lagrangian density $\mathcal{L}^{(m)}$ does not depend on $\phi$, but the scalar field is directly coupled to the Ricci scalar $R$. The scalar field potential $V(\phi)$ constitutes a natural generalization of the cosmological constant ~\cite{Capozziello_book}.\\
From the action of ST theories of gravity, the BD's action can be gotten by ~\cite{faraoni_book}
\begin{equation}
	\label{eq:STaBD}
	f(\phi)=\frac{\phi}{8\pi}, \qquad\omega(\phi)=\frac{\omega_{0}}{8\pi\phi}
\end{equation}
where $\omega_{0}$ is a constant, and the potential is rescaled by a factor $16\pi$.\\
The ST field equations can be obtained from a variational principle. The variation of the action (\ref{eq:accionST}) with respect to $\delta g^{ab}$ gives
\begin{eqnarray}
	\label{eq:varST}
	\fl\delta S^{(ST)}=\int_{\mathcal{M}}d^{4}x\delta(\sqrt{-g})\Big[
	\frac{f(\phi)}{2}
	R-\frac{\omega(\phi)}{2}g^{cd}\nabla_{c}\phi\nabla_{d}\phi-V(\phi)\Big]
	\nonumber\\
	+\int_{\mathcal{M}}d^{4}x\sqrt{-g}\Big[
	\frac{f(\phi)}{2}\delta R-\frac{\omega(\phi)}{2}\delta(
	g^{cd})\nabla_{c}\phi\nabla_{d}\phi\Big] +\delta S^ {(m)}.
\end{eqnarray}
Taking into account the equation (\ref{eq:deltaR}), we get
\begin{eqnarray}
	\label{eq:varST1}
\fl	\delta
	S^{(ST)}=
	\int_{\mathcal{M}}d^{4}x\sqrt{-g}\left[\frac{f(\phi)}{2}\left(
	R_{ab}-\frac{1}{2}g_{ab}R\right)+\frac{1}{2}g_{ab}V(\phi)\right.\nonumber\\
	\left.-
	\frac{\omega(\phi)}{2}\left(\nabla_{a}\phi\nabla_{b}\phi-\frac{1}{2}g_{ab}g^{cd
	}\nabla_{c} \phi\nabla_{d}\phi\right)\right]
	\delta
	g^{ab}\nonumber\\
	+\int_{\mathcal{M}}d^{4}x\sqrt{-g}\frac{f(\phi)}{2}\left[\nabla_{c}\left(g^{
		ab } \delta\christoffel{c}{ab}\right)-\nabla_{b}\left(g^{ab}
	\delta\christoffel{c}{ac}\right)\right]+\delta S^ {(m)}.
\end{eqnarray}
Let us write the second integral in the following way
\begin{equation}
	\label{eq:varBDb1}
	\delta
	S^{(ST)}_{B}=\int_{\mathcal{M}}d^{4}x\sqrt{-g}\frac{f(\phi)}{2}\nabla_{d}\left(
	g^{ab}\delta\christoffel{d}{ab}-g^{ad}\delta\christoffel{c}{ac}\right).
\end{equation}
The term in parentheses is given by (e.g. see ~\cite{Alejo})
\begin{equation}
	\label{eq:deltaschristoffel}
	g^{ab}\delta\christoffel{d}{ab}-g^{ad}\delta\christoffel{c}{ac}=g_{ef}\nabla^{d}
	\delta g^{ef}-\nabla_{c}\delta g^{dc}.
\end{equation}
Using the above relation and the fact about the metric
%Bajo confirmación
compatibility ($\nabla_{c}g_{ab}=0$), the term (\ref{eq:varBDb1}) yields
\begin{equation}
	\label{eq:varBDb2}
	\delta
	S^{(ST)}_{B}=\int_{\mathcal{M}}d^{4}x\sqrt{-g}\frac{f(\phi)}{2}\left(
	g_{ef}\Box\delta g^{ef}-\nabla_{e}\nabla_{f}\delta g^{ef} \right),
\end{equation}
where the D'Alembert operator definition has been used, i.e. $\Box\equiv\nabla_{d}\nabla^{d}$. It allow us to define the next quantities to express the integral above in a different way
%Bajo confirmación

\begin{equation}
	\label{eq:M}
	M_{c}=\frac{f(\phi)}{2} g_{ef}\nabla_{c}(\delta g^{ef})-\frac{1}{2}(\delta g^{ef})g_{ef}\nabla_{c}f(\phi)
\end{equation}
y
\begin{equation}
	\label{eq:N}
	N^{c}=\frac{f(\phi)}{2}\nabla_{f}(\delta g^{cf})-\frac{1}{2}(\delta g^{cf})\nabla_{f}f(\phi).
\end{equation}
The quantities $M_{c}$ and $N^{c}$ allow us to write the equation (\ref{eq:varBDb2}) as (for details view \ref{AppendixA})
\begin{equation}
	\label{eq:varBDb3}
	\fl\delta S^{(ST)}_{B}=\frac{1}{2}\int_{\mathcal{M}}d^{4}x\sqrt{-g}\delta
	g^{ef}(g_{ef}\Box f(\phi)-\nabla_{e}\nabla_{f}f(\phi))+\int_{\mathcal{M}
	}d^{4}x\sqrt{-g}(\nabla^{c}M_{c}-\nabla_{c}N^{c}).
\end{equation}
Thus, the variation of the action (\ref{eq:varST1}) takes the form
	\begin{eqnarray}\label{eq:varaccionSTconNyM}
		\fl\delta
		S^{(ST)}=\int_{\mathcal{M}}d^{4}x\sqrt{-g}\Bigg[\frac{f(\phi)}{2}\left(
		R_{ab}-\frac{1}{2}g_{ab}R\right)+\frac{1}{2}g_{ab}V(\phi)\nonumber\\
-		\frac{\omega(\phi)}{2}\left(\nabla_{a}\phi\nabla_{b}\phi-\frac{1}{2}g_{ab}g^{cd
		}\nabla_{c} \phi\nabla_{d}\phi\right)+\frac{1}{2}(g_{ab}\Box f(\phi)-\nabla_{a}\nabla_{b}f(\phi))\Bigg]\delta g^{ab}\nonumber\\
		+\oint_{\partial\mathcal{M}} d^{3}y\sqrt{|h|}\epsilon n^{c}M_{c}-\oint_{\partial\mathcal{M}} d^{3}y\sqrt{|h|}\epsilon n_{c}N^{c}+\delta S^ {(m)},
	\end{eqnarray}
where the Gauss-Stokes theorem  (\ref{eq:gauss-stokes}) has been used in the boundary term. Evaluating the terms $M_{c}$ and $N^{c}$ at the boundary, we have
\begin{eqnarray}
	\label{eq:M1}
	M_{c}\Big\vert_{\partial\mathcal{M}}&=\frac{f(\phi)}{2} g_{ef}\nabla_{c}\delta g^{ef}=-\frac{f(\phi)}{2}\delta^{a}_{f}g^{bf}\partial_{c}\delta g_{ab}\nonumber\\
	&=-\frac{f(\phi)}{2} g^{ba}\partial_{c}\delta g_{ab}
\end{eqnarray}
and
\begin{equation}
	\label{eq:N1}
	N^{c}\Big\vert_{\partial\mathcal{M}}=-\frac{f(\phi)}{2} g^{ac}g^{bf}\partial_{f}\delta g_{ab}.
\end{equation}
Using (\ref{eq:metricafrontera}) we compute the following terms  that appear in the integrals (\ref{eq:varaccionSTconNyM})
\begin{eqnarray}
	\label{eq:M3}
	n^{c}M_{c}\Big\vert_{\partial\mathcal{M}}&=-\frac{f(\phi)}{2} n^{c}(h^{ab}+\epsilon n^{a}n^{b})\partial_{c}\delta g_{ab}\nonumber\\
	&=-\frac{f(\phi)}{2} n^{c}h^{ab}\partial_{c}\delta g_{ab}
\end{eqnarray}
and
\begin{eqnarray}
	\label{eq:N2}
	n_{c}N^{c}\Big\vert_{\partial\mathcal{M}}&=-\frac{f(\phi)}{2} n_{c}(h^{ac}+\epsilon n^{a}n^{c})(h^{bf}+\epsilon n^{b}n^{f})\partial_{f}(\delta g_{ab})\nonumber\\
	&=-\frac{f(\phi)}{2} n^{a}h^{bf}\partial_{f}(\delta g_{ab})=0,
\end{eqnarray}
where we have used the facts that $n_{c}h^{ac}=0$, $\epsilon^{2}=1$ and the tangential derivative $h^{bf}\partial_{f}(\delta g_{ab})$ to vanish (e.g., see ~\cite{Poisson_book}).\\
The variation of the action (\ref{eq:varaccionSTconNyM}) takes the form
	\begin{eqnarray}\label{eq:varBD3}
		\fl\delta
		S^{(ST)}=\int_{\mathcal{M}}d^{4}x\sqrt{-g}\Bigg[\frac{f(\phi)}{2}\left(
		R_{ab}-\frac{1}{2}g_{ab}R\right)+\frac{1}{2}g_{ab}V(\phi)\nonumber\\-
		\frac{\omega(\phi)}{2}\left(\nabla_{a}\phi\nabla_{b}\phi-\frac{1}{2}g_{ab}g^{cd
		}\nabla_{c} \phi\nabla_{d}\phi\right)
		+\frac{1}{2}(g_{ab}\Box f(\phi)-\nabla_{a}\nabla_{b}f(\phi))\Bigg]\delta g^{ab}\nonumber\\
		-\frac{1}{2}\oint_{\partial\mathcal{M}}d^{3}y\sqrt{|h|}\epsilon f(\phi) n^{c}h^{ab}\partial_{c}(\delta g_{ab})+\delta S^ {(m)}.
	\end{eqnarray}
As previously mentioned for GR, the last integral can be vanished arguing that, in addition to the variation of the metric $\delta g^{ab}$, its first derivative $\partial_{c}\delta g_{ab}$ to vanish in the bpundary. Instead of, we use the boundary term type GYH for ST theories ~\cite{Padilla,Dyer}
\begin{equation}
	\label{eq:GYH}
	S^{(ST)}_{GYH}=2\oint_{\partial\mathcal{M}}d^{3}y\sqrt{|h|}\epsilon \frac{f(\phi)}{2} K.
\end{equation}
The variation of this term with respect to $\delta g^{ab}$ is
\begin{equation}
	\label{eq:GYH1ST}
	\delta S^{(ST)}_{GYH}=\oint_{\partial\mathcal{M}}d^{3}y\sqrt{|h|}\epsilon f(\phi)\delta K.
\end{equation}
Taking into account (\ref{eq:vartrazak}), the above equation gives
\begin{equation}
	\label{eq:GYH2}
	\delta S^{(ST)}_{GYH}=\frac{1}{2}\oint_{\partial\mathcal{M}}d^{3}y\sqrt{|h|}\epsilon f(\phi) n^{c}h^{ab}\partial_{c}\delta g_{ab}.
\end{equation}
Thus, we can see that the term type GYH cancels with the second integral of the equation (\ref{eq:varBD3}).\\
Finally, using (\ref{eq:varaccionM}), the variation of the action of ST theories yields
\begin{eqnarray}\label{eq:varST4}
	\fl\delta
	S^{(ST)}=
	\int_{\mathcal{M}}d^{4}x\sqrt{-g}\Bigg[\frac{f(\phi)}{2}\left(
	R_{ab}-\frac{1}{2}g_{ab}R\right)+\frac{1}{2}g_{ab}V(\phi)\nonumber\\
	-
	\frac{\omega(\phi)}{2}\left(\nabla_{a}\phi\nabla_{b}\phi-\frac{1}{2}g_{ab}g^{cd
	}\nabla_{c} \phi\nabla_{d}\phi\right)
	+\frac{1}{2}(g_{ab}\Box f(\phi)-\nabla_{a}\nabla_{b}f(\phi))\nonumber\\
	\hspace{1.5cm}-\frac{1}{2}T^{(m)}_{ab}\Bigg]\delta g^{ab}.
\end{eqnarray}
Imposing that this variation becomes stationary
\begin{equation}
	\label{eq:varSTe}
	\frac{1}{\sqrt{-g}}\frac{\delta S^{(ST)}}{\delta g^{ab}}=0,
\end{equation}
we get
\begin{equation}
	\label{eq:campoST1}
	\fl f(\phi)G_{ab}=T^{(m)}_{ab}+\omega(\phi)(\nabla_{a}\phi\nabla_{b}\phi-\frac{1}{2}g_{ab}\nabla^{c}\phi\nabla_{c}\phi)
+(\nabla_{a}\nabla_{b}f(\phi)- g_{ab}\Box f(\phi))-g_{ab}V(\phi),
\end{equation}
which are the field equations in the metric formalism of ST theories of gravity.\\
Since the action (\ref{eq:accionST}) it depends on the metric as the scalar field $\phi$, the variation of the action (\ref{eq:accionST}) with respect to $\delta\phi$ is calculated
\begin{equation}\label{eq:varSTphi}
	\delta S^{(ST)}=
	\int_{\mathcal{M}}d^{4}x\sqrt{-g}\left[\frac{1}{2}R\delta f(\phi)-\frac{1}{2}\delta(\omega(\phi)\nabla^{c}\phi
	\nabla_{c}\phi)-\delta V(\phi)\right].
\end{equation}
Allow us to write, $\delta f(\phi)=\frac{df(\phi)}{d\phi}\delta\phi=f_{\phi}\delta\phi$.\\
Now, the second term in the integral we can write it as
\begin{eqnarray}
	\label{eq:varomegaphi}
	\delta(\omega(\phi)\nabla^{c}\phi\nabla_{c}\phi)&=\nabla^{c}\phi\nabla_{c}\phi\delta\omega(\phi)+
	\omega(\phi)\delta(\nabla^{c}\phi\nabla^{c}\phi)\nonumber\\
	&=\nabla^{c}\phi\nabla_{c}\phi\omega_{\phi}\delta\phi+2\omega(\phi)\nabla^{c}\phi\nabla_{c}\delta\phi.
\end{eqnarray}
Thus, the variation gives
\begin{equation}\label{eq:varSTphi1}
\fl	\delta S^{(ST)}=\int_{\mathcal{M}}d^{4}x\sqrt{-g}\left[\frac{1}{2}R f_{\phi}-\frac{1}{2}\omega_{\phi}\nabla^{c}\phi
	\nabla_{c}\phi-V_{\phi}\right]\delta\phi-\int_{\mathcal{M}}d^{4}x\sqrt{-g}\omega(\phi)\nabla^{c}\phi\nabla_{c}\delta\phi.
\end{equation}
we define the following quantity for can be expressed diferently the above integral
\begin{equation}
	\label{eq:L}
	L^{c}=\omega(\phi)\nabla^{c}\phi\delta\phi.
\end{equation}
The covariant derivative of $L^{c}$ is
\begin{eqnarray*}
	\nabla_{c}L^{c}&=\nabla_{c}(\omega(\phi))\nabla^{c}\phi\delta\phi+
	\omega(\phi)\nabla_{c}(\nabla^{c}\phi\delta\phi)\\
	&=\omega_{\phi}\nabla_{c}\phi\nabla^{c}\phi\delta\phi+ \omega(\phi)\nabla_{c}(\nabla^{c}\phi\delta\phi).
\end{eqnarray*}
Because
\begin{equation}
	\label{eq:nablaphi1}
	\nabla^{c}\phi\nabla_{c}(\delta\phi)=\nabla_{c}\left(\delta\phi\nabla^{c}\phi\right)-\delta\phi\Box\phi,
\end{equation}
the second term in (\ref{eq:varSTphi1}) takes the form
\begin{equation}\label{eq:varSTphi3}
	\fl\delta S^{(ST)}=\int_{\mathcal{M}}d^{4}x\sqrt{-g}\Big[\frac{1}{2}R f_{\phi}+\frac{1}{2}\omega_{\phi}\nabla^{c}\phi
	\nabla_{c}\phi+\omega(\phi)\Box\phi- V_{\phi}\Big]\delta\phi-\int_{\mathcal{M}}d^{4}x\sqrt{-g}\nabla_{c}L^{c}.
\end{equation}
Using the Gauss-Stokes theorem (\ref{eq:gauss-stokes}) at the divergence term, we have
\begin{equation}
	\label{eq:L1}
	\fl\int_{\mathcal{M}}d^{4}x\sqrt{-g}\nabla_{c}L^{c}=\oint_{\partial\mathcal{M}}d^{3}y\sqrt{|h|}\epsilon n_{c}L^{c}=\oint_{\partial\mathcal{M}}d^{3}y\sqrt{|h|}\epsilon n_{c}\omega(\phi)\nabla^{c}\phi\delta\phi.
\end{equation}
Imposing that the variation of the scalar field in the boundary vanishes
\begin{equation}\label{eq:imposdeltaphi}
	\delta\phi\Big\vert_{\partial\mathcal{M}}=0,
\end{equation}
we can see that the Gauss-Stokes term cancels-off.\\
Now, the variation of the term type GYH for ST theories(\ref{eq:GYH}) with respect to $\delta\phi$ yields
\begin{equation}
	\label{eq:GYH1}
	\delta S^{(ST)}_{GYH}=\oint_{\partial\mathcal{M}}d^{3}y\sqrt{|h|}\epsilon f_{\phi} K\delta\phi,
\end{equation}
because the imposition (\ref{eq:imposdeltaphi}), the above term vanishes.
Wherewith, the variation of the action (\ref{eq:varSTphi3}) gives
\begin{equation}\label{eq:varSTphi4}
	\delta S^{(ST)}=\int_{\mathcal{M}}d^{4}x\sqrt{-g}\left[\frac{1}{2}R f_{\phi}+\frac{1}{2}\omega_{\phi}\nabla^{c}\phi
	\nabla_{c}\phi+\omega(\phi)\Box\phi- V_{\phi}\right]\delta\phi.
\end{equation}
Imposing that this variation become stationary
\begin{equation} \label{eq:varSTe1}
	\frac{1}{\sqrt{-g}}\frac{\delta S^{(ST)}}{\delta\phi}=0,
\end{equation}
we have
\begin{equation}\label{eq:campoST2}
	\omega(\phi)\Box\phi+\frac{1}{2}R f_{\phi}+\frac{1}{2}\omega_{\phi}\nabla^{c}\phi
	\nabla_{c}\phi-V_{\phi}=0.
\end{equation}
which are the field equation for the scalar field in ST theories of gravity.
%=================================================================================================================================================
\subsection{Field Equations in $f(R)$ theories}
As a natural extension of GR and higher order theories, $f(R)$ theories emerge, which consider an arbitrary function of the Ricci scalar.\\
The action  $f(R)$ is~\cite{DeFelice}
\begin{equation}\label{eq:accionf(R)}
	S^{f(R)}=\frac{1}{2}\int_{\mathcal{M}}d^{4}x\sqrt{-g}f(R)+S^{(m)},
\end{equation}
where $f(R)$ is a non-linear analytical function of the Ricci scalar and $S^{(m)}$ is given by (\ref{eq:accionM}). In the paper~\cite{Alejo}, shows how the field equations are obtained taking into account the boundary term type GYH for $f(R)$. Here show the main results found there.\\
The variation of the action with respect to $\delta g^{ab}$ is
\begin{eqnarray}
\fl\delta S^{f(R)}=
		\frac{1}{2}\int_{\mathcal{M}}d^{4}x\sqrt{-g}\left(f_{R}R_{ab}-\frac{1}{2}g_{ab}f(R)+g_{ab}\Box f_{R}-\nabla_{a}\nabla_{b}f_{R}\right)\delta g^{ab}\nonumber\\
		+\frac{1}{2}\int_{\mathcal{M}}d^{4}x(\nabla^{c}H_{c}-\nabla_{c}I^{c})+\delta S^{(m)},
\end{eqnarray}
where the terms $H_{c}$ and $I^{c}$ are given by
\begin{equation}
	H_{c}=f_{R}g_{ab}\nabla_{c}\delta g^{ab}-\delta g^{ab}g_{ab}\nabla_{c}f_{R}
\end{equation}
and
\begin{equation}
	I^{c}=f_{R}\nabla_{e}\delta g^{ce}-\delta g^{ce}\nabla_{e}f_{R}.
\end{equation}
Here $f_{R}=\frac{df}{dR}$. Using the Gauss-Stokes theorem to the divergence term in the variation and evaluating the terms $n^{c}H_{c}$ and $n_{c}I^{c}$ at the boundary, we have
	\begin{eqnarray}\label{eq:varaccionf(R)2}
		\fl\delta S^{f(R)}=
		\int_{\mathcal{M}}d^{4}x\sqrt{-g}\left(f_{R}R_{ab}-\frac{1}{2}g_{ab}f(R)+g_{ab}\Box f_{R}-\nabla_{a}\nabla_{b}f_{R}\right)\delta g^{ab}\nonumber\\
		-\oint_{\partial\mathcal{M}}d^{3}y
		\sqrt{|h|}\epsilon f_{R}n^{c}h^{ab}\partial_{c}\delta g_{ab}+\delta S^{(m)}.
	\end{eqnarray}
The boundary term type GYH for $f(R)$ is ~\cite{Dyer}
\begin{equation}
	S^{f(R)}_{GYH}=\oint_{\partial\mathcal{M}}d^{3}y\sqrt{|h|}\epsilon f_{R}K,
\end{equation}
The variation of the above action gives
\begin{equation}
	\delta S^{f(R)}_{GYH}=\oint_{\partial\mathcal{M}}d^{3}y\sqrt{|h|}\epsilon f_{RR}K\delta R+\frac{1}{2}\oint_{\partial\mathcal{M}}d^{3}y\sqrt{|h|}\epsilon n^{d} f_{R} h^{ab}\partial_{d}\delta g_{ba}.
\end{equation}
The second term of the above equation cancels the boundary term of the equation (\ref{eq:varaccionf(R)2}), but in addition needs to impose $\delta R=0$ in the boundary to obtain the field equations~\cite{Alejo, Saltas}.\\
%We will see in more detail the equivalences between theories of gravity ST and $f(R)$ in the next section, but it should be noted that for $f(R)$ an additional imposition was made on $\delta R$, which in the equivalence is $\delta\phi$ in ST.\\
Taking into account the variation of the matter action  (\ref{eq:varaccionM}) and imposing that the variation for $f(R)$ theories becomes stationary
\begin{equation}
	\frac{1}{\sqrt{-g}}\frac{\delta S^{f(R)}}{\delta g^{ab}}=0,
\end{equation}
thus, we have
\begin{equation}\label{eq:campof(R)}
	f_{R}R_{ab}-\frac{1}{2}g_{ab}f(R)+g_{ab}\Box f_{R}-\nabla_{a}\nabla_{b}f_{R}=T^{(m)}_{ab}.
\end{equation}
which are the field equations for $f(R)$ theories.\\
In this section we recover in the variational approach the set of field equations for GR, ST and $f(R)$ theories emphasizing the boundary problem. We explore directly the equivalence between ST and $f(R)$ theories at the GYH boundary term, and it is clear that the boundary term makes the theory well defined mathematical problem. It is important to notice that in the literature the equivalence problem has been widely studied ~\cite{Belenchia, Capone, Ntahompagaze, Sotiriou-1,Odinstov,Nojiri}, but  in this paper it was shown how the field equations were obtained for ST theories with the GYH boundary term, in complete agreement with previous work ~\cite{Nojiri,Nojiri_2, Odinstov, Saltas}, but conecting a previous work \cite{Alejo} through the equivalence in the important issue of the boundary for both theories. Also, the condition to get the equation to $\phi$, it had to be imposed on the boundary that the variation $\delta\phi$ be equal to zero. The variational approach in $f(R)$ gravity brings the condition $\delta R=0$ at the boundary in total agreement with the equivalence between both theories, showing the mathematical power of the equivalence.\\ 
A more detailed analysis of the equivalences will be discussed in the next section.
%=================================================================================================================================================
\section{\label{sec::equivalencias}Equivalence between ST and $f(R)$ theories}
The equivalence between ST and $f(R)$ theories has been broadly studied at the classical level, e.g., in ~\cite{Belenchia, Capone, Ntahompagaze, Sotiriou-1,Odinstov,Nojiri}, but also a quantum level ~\cite{Ohta, Ruf}. In this paper shows the equivalence between the actions and the field equations, but as we will see in the next section, in addition we will show them in the cosmological perturbations.\\
We start from the following ST action without a kinetic term in the scalar field
\begin{equation}\label{eq:accioncampoaux}
	S=\int_{\mathcal{M}}d^{4}x\sqrt{-g}(\psi(\phi)R-V(\phi)),
\end{equation} 
donde $\phi$ has been included as an auxiliary field. \\
when $f_{\phi\phi}\neq0$ in the above action, we can set
\begin{eqnarray}\label{eq:psienphi}
	\psi&=f_{\phi}\\\label{eq:v(phi)enR}
	V(\phi)&=\phi f_{\phi}-f(\phi)=\phi\psi(\phi)-f(\phi),
\end{eqnarray}
Thus, the action (\ref{eq:accioncampoaux}) takes the form
\begin{equation}\label{eq:accioncampoaux2}
	S=\int_{\mathcal{M}}d^{4}x\sqrt{-g}(f_\phi(R-\phi)+f(\phi)).
\end{equation} 
If $\phi=R$, we have
\begin{equation}\label{eq:equivf(R)}
	\psi=f_{R}
\end{equation} 
and we recover the action (\ref{eq:accionf(R)}). Moreover, the variation with respect to $\phi$ of the above action gives
\begin{equation}
	f_{\phi\phi}(R-\phi)=0,
\end{equation}
if $f_{\phi\phi}\neq0$ it implies that
\begin{equation}\label{eq:equivphi}
	\phi=R. 
\end{equation}
The action (\ref{eq:accioncampoaux}) corresponds to the action  (\ref{eq:accionST}) of ST theories with the parameter $\omega(\phi)=0$.\\
If we start with the field equations $f(R)$ (rewriting the equations (\ref{eq:campof(R)}) for to include the Einstein tensor $G_{ab}$)
\begin{equation}
	G_{ab}f_{R}=T^{(m)}_{ab}+\nabla_{a}\nabla_{b}f_{R}-g_{ab}\Box f_{R}+\frac{1}{2}g_{ab}(f-Rf_{R}),
\end{equation}
Taking into account (\ref{eq:equivf(R)}) in the above field equations, we get
\begin{eqnarray*}
	G_{ab}\psi&=T^{(m)}_{ab}+\nabla_{a}\nabla_{b}\psi-g_{ab}\Box\psi+\frac{1}{2}g_{ab}(f(\phi)-\phi\psi)\\
	&=T_{ab}+\nabla_{a}\nabla_{b}\psi-g_{ab}\Box\psi-g_{ab}V(\phi),
\end{eqnarray*}
where it has been used (\ref{eq:v(phi)enR}), with the potential rescaled by $\frac{1}{2}$. The above equations are the field equations  (\ref{eq:campoST1}) for ST theories with the parameter $\omega(\phi)=0$.
%===================================================================================================================================================================================================================================================================================
\section{\label{sec::perturbaciones}Cosmological Perturbations}
In this section we study the Friedmann equations in a homogeneous and isotropic universe with the metric Friedmann-Lemaître-Robertson-Walker (FLRW) as the background metric for the ST and $f(R)$ theories. Then we calculate the linear cosmological perturbations under conformal-Newtonian gauge for the theories above mentioned. Note that the equations found by $f(R)$ theories for both the background and the perturbed ones were found under the equivalence relations with the ST theories.
%===========================================================================
\subsection{Background Universe}
Consider a statistically spatially homogeneous and isotropic universe with the spatially flat FLRW metric as background
\begin{equation}\label{eq:FLRWconf}
	ds^{2}=a^{2}(\eta)(-d\eta^{2}+\delta_{\mu\nu}dx^{\mu}dx^{\nu}).
\end{equation}
The energy conservation is
\begin{equation}\label{eq:consenergia}
	\nabla^{b}T_{ab}=0,
\end{equation}
where
\begin{equation}\label{eq:m-e}
	T_{ab}=pg_{ab}+(p+\rho)u_{a}u_{b}
\end{equation}
is the stress-energy tensor for perfect fluid. With this, the energy conservation gives
\begin{equation}\label{eq:consmateria}
	\dot{\rho}+3H(p+\rho)=0.
\end{equation}
Here, $p$ is the fluid pressure, $\rho$ the energy density y $u^{a}$ is the four-velocity of the fundamental observers.\\
The Friedmann equations for the evolution of the background in ST theories are ~\cite{faraoni_book}
\begin{equation}\label{eq:friedmannST1conf}
	3\mathcal{H}^{2}f=\rho a^{2}+\frac{\omega}{2}\phi'^{2}+Va^{2}-3\mathcal{H}f'
\end{equation}
and
\begin{equation}\label{eq:friedmannSTconf2}
	-(2\mathcal{H}'+\mathcal{H}^{2})f=pa^{2}+\frac{1}{2}\omega \phi'^{2}+\mathcal{H}f'+f''-Va^{2},
\end{equation}
where $\mathcal{H}\equiv\frac{\dot{a}(\eta)}{a(\eta)}$. 
The equation for the evolution of the scalar field is
\begin{equation}\label{eq:friedmann3conf}
	\omega(\phi''+2\mathcal{H}\phi')=3f_{\phi}(\mathcal{H}'+\mathcal{H}^{2})-\frac{1}{2}\omega_{\phi}\phi'^{2}-V_{\phi}a^{2}.
\end{equation}
To obtain the Friedmann equations for BD theory, must be taking into account the relations (\ref{eq:STaBD}) in the friedmann equation for ST theories.\\
From the equivalence relation (\ref{eq:equivf(R)}), we have
\begin{eqnarray}\label{eq:f(rr)}
	\psi_{\phi}&=f_{RR}\\
	\psi_{\phi\phi}&=f^{(3)}_{R},
\end{eqnarray}
where $f^{(3)}_{R}=\frac{d^{3}f}{dR^{3}}$. Replacing (\ref{eq:v(phi)enR}) and the above relations in the equations (\ref{eq:friedmannST1conf}) and (\ref{eq:friedmannSTconf2}) with the parameter $\omega(\phi)=0$, we come to Friedmann equations for the $f(R)$ theories
\begin{equation}\label{eq:friedmannf(R)conf1}
	3\mathcal{H}^{2}f_{R}=\rho a^{2}+\frac{a^{2}}{2}(Rf_{R}-f(R))-3\mathcal{H}f_{RR}R'
\end{equation}
and
\begin{equation}\label{eq:friedmannf(R)conf2}
	\fl-(2\mathcal{H}'+\mathcal{H}^{2})f_{R}=pa^{2}+\mathcal{H}R'f_{RR}+\frac{a^{2}}{2}(f(R)-Rf_{R})+R''f_{RR}+R'^{2}f^{(3)}_{R}.
\end{equation}
As mentioned above, one of the motivations for MG theories, is to explain the accelerating expansion of the universe. For ST theories, given a potential $V(\phi)$ ~\cite{Hu-Sawicki,Yasunori,Gannouji} we can get a universe in accelerating expansion, while for $f(R)$ theories, the same function is responsible for achieve it ~\cite{Clifton,Morita,Bahamonde}.\\
Through the equivalence we have found the Friedmann equations for the theories $f(R)$ starting from the ST equations, taking the parameter $\omega=0$. Next we will find the Friedmann equations perturbed for both theories in a complete general framework.
%===============================================================================================================================================
\subsection{Equivalence between Cosmological Perturbations in ST and $f(R)$ gravity}
The line element of the perturbed universe is
\begin{equation}\label{eq:lineapert}
\fl ds^{2}=a^{2}(\eta)[-(1+2A)d\eta^{2}-2B_{\mu}d\eta dx^{\mu}+[(1-2D)\delta_{\mu\nu}+2E_{\mu\nu}]dx^{\mu}dx^{\nu}],
\end{equation}
where $A$, $B_{\mu}$, $D$ and $E_{\mu\nu}$ are metric perturbations. Now, we can descomposed the $0-i$ and the $i-j$ components of the metric tensor into, scalar, vector and tensor parts
\begin{equation}
B_{\mu}=-B_{\mu}+B_{\mu}^{V}, \quad where\quad \delta^{\mu\nu}B_{\mu}^{V}=0
\end{equation}
and
\begin{equation}
E_{\mu\nu}=E_{\mu\nu}^{S}+E_{\mu\nu}^{V}+E_{\mu\nu}^{T},
\end{equation}
here
\begin{eqnarray}
E_{\mu\nu}^{S}&=\left(\partial_{\mu}\partial_{\nu}-\frac{1}{3}\delta_{\mu\nu}\nabla^{2}\right)E,\\
E_{\mu\nu}^{V}&=-\frac{1}{2}(\nabla_{\mu}E_{\nu}+\nabla_{\nu}E_{\mu}),\quad where\quad \delta^{\mu\nu}E_{\mu},_{\nu}=0,\\
&\delta^{\mu\kappa}E_{\mu\nu}^{T},_{\kappa}=0,\qquad\delta^{\mu\nu}E_{\mu\nu}^{T}=0.
\end{eqnarray}
Due to this division perturbation to fisrt order, we can study the scalar, vector and tensor perturbations separately. In the following we show the perturbations for ST theories and the equivalences with $f(R)$ gravity into the components above mentioned
\subsubsection{Scalar Perturbations}
Scalar metric perturbations  are describes by the line element  ~\cite{Steinhardt_book}
\begin{equation}\label{eq:lineapertesc2}
\fl ds^{2}=a^{2}(\eta)\Big[-(1+2A)d\eta^{2}+2B_{,\mu}d\eta dx^{\mu}
+\left[(1-2\psi)\delta_{\mu\nu}+2E_{,\mu\nu}\right]dx^{\mu}dx^{\nu}\Big].
\end{equation}
where $A$, $B$, $\psi$ and $E$ are scalar perturbations. The curvature perturbation $\psi$ is defined by $\psi\equiv D+\frac{1}{3}\nabla^{2}E$. To find the linear perturbations of ST theories, the field equations (\ref{eq:campoST1}) are perturbed, taking into account the metric (\ref{eq:lineapertesc2}). Here, $\delta\phi$ represents the perturbation of the scalar field. The perturbed Friedmann equations are 
\begin{eqnarray}\label{eq:pertST00}
&\fl\big[-2\nabla^{2}\psi+6\mathcal{H}\psi'+6\mathcal{H}^{2}A+2\mathcal{H}\nabla^{2}B-2\mathcal{H}\nabla^{2}E'\big]\bar{f}(\phi)-3\mathcal{H}^{2}\bar{f}_{\phi}\delta\phi=-a^{2}\delta\rho\nonumber\\
&\fl+\bar{\omega}(\phi)\left(\bar{\phi}'^{2}A-\bar{\phi}'\delta\phi'\right)-\frac{1}{2}\bar{\omega}_{\phi}\bar{\phi}'^{2}\delta\phi+3\mathcal{H}(\bar{f}_{\phi}\delta\phi)'-6\mathcal{H}\bar{f}_{\phi}\bar{\phi}'A-\bar{f}_{\phi}\partial^{2}\delta\phi\nonumber\\
&\fl-\bar{f}_{\phi}\bar{\phi}'(\partial^{2}B+3\psi'-\partial^{2}E')-a^{2}\bar{V}_{\phi}\delta\phi,
\end{eqnarray}
which is the $0-0$ perurbed component, 
\begin{eqnarray}\label{eq:pertSTte}
&\fl-2(\psi'+\mathcal{H}A)_{,\mu}\bar{f}(\phi)=-a^{2}(\bar{\rho}+\bar{p})(v_{,\mu}-B_{,\mu})-\bar{\omega}(\phi)\bar{\phi}'\partial_{\mu}\delta\phi-\bar{f}_{\phi}\partial_{\mu}\delta\phi'\nonumber\\
&\fl-\bar{f}_{\phi\phi}\bar{\phi}'\partial_{\mu}\delta\phi+A_{,\mu}\bar{f}_{\phi}\bar{\phi}'+\mathcal{H}\bar{f}_{\phi}\partial_{\mu}\delta\phi,
\end{eqnarray}
which is the $0-\mu$ perurbed component, 
\begin{eqnarray}\label{eq:pertSTet}
&\fl2(\psi'+\mathcal{H}A-\mathcal{H}'B+\mathcal{H}^{2}B)_{,\mu}\bar{f}(\phi)=a^{2}(\bar{\rho}+\bar{p})v_{,\mu}+\bar{\omega}(\phi)[(\bar{\phi}')^{2}B_{,\mu}+\bar{\phi}'\partial_{\mu}\delta\phi]
\nonumber\\
&\fl+B_{,\mu}(\bar{f}_{\phi}\bar{\phi}')'-2\mathcal{H}B_{,\mu}\bar{f}_{\phi}\bar{\phi}'+\bar{f}_{\phi}\partial_{\mu}\delta\phi'+\bar{f}_{\phi\phi}\bar{\phi}'\partial_{\mu}\delta\phi-A_{,\mu}\bar{f}_{\phi}\bar{\phi}'-\mathcal{H}\bar{f}_{\phi}\partial_{\mu}\delta\phi,
\end{eqnarray}
which is the $\mu-0$ perturbed component, 
\begin{eqnarray}\label{eq:pertSTee}
&\fl-(2\mathcal{H}'+\mathcal{H}^{2})\bar{f}_{\phi}\delta\phi\delta_{\mu\nu}+[2\psi''-\nabla^{2}(\psi-A)+\mathcal{H}(2A'+4\psi')+(4\mathcal{H}'+2\mathcal{H}^{2})A\nonumber\\
&\fl+\nabla^{2}B'+2\mathcal{H}\nabla^{2}B]\bar{f}(\phi)\delta_{\mu\nu}-(\nabla^{2}E''+2\mathcal{H}\nabla^{2}E')\bar{f}(\phi)\delta_{\mu\nu}\nonumber\\
&\fl+(\psi-A-B'-2\mathcal{H}B+E''+2\mathcal{H}E'),_{\mu\nu}\bar{f}(\phi)=\delta pa^{2}\delta_{\mu\nu}+\bar{p}a^{2}\left(\Pi,_{\mu\nu}-\frac{1}{3}\delta_{\mu\nu}\nabla^{2}\Pi\right)\nonumber\\
&\fl+\bar{\omega}(\phi)\left(\bar{\phi}'\delta\phi'-\bar{\phi}'^{2}A\right)\delta_{\mu\nu}+\frac{1}{2}\bar{\omega}_{\phi}\bar{\phi}'^{2}\delta\phi\delta_{\mu\nu}+\bar{f}_{\phi}\left(\partial_{\mu}\partial_{\nu}-\delta_{\mu\nu}\partial^{2}\right)\delta\phi\nonumber\\
&\fl+\mathcal{H}\left((\bar{f}_{\phi}\delta\phi)'-2\bar{f}_{\phi}\bar{\phi}'A\right)\delta_{\mu\nu}+\bar{f}_{\phi}\bar{\phi}'\left(B_{,\mu\nu}-2\psi'\delta_{\mu\nu}-E'_{,\mu\nu}\right)+(\bar{f}_{\phi}\delta\phi)''\delta_{\mu\nu}\nonumber\\
&\fl-2(\bar{f}_{\phi}\bar{\phi}')'A\delta_{\mu\nu}+\bar{f}_{\phi}\bar{\phi}'\left(\partial^{2}E'-\partial^{2}B-A'\right)\delta_{\mu\nu}-a^{2}\bar{V}_{\phi}\delta\phi\delta_{\mu\nu},
\end{eqnarray}
and finally, the $\mu-\nu$ perturbed component. In order to find the relationship between the scalar potentials and anisotropic pressure, we take the off-diagonal part, after having calculated the trace of the above equation
\begin{equation}\label{eq:relpotsgeneralconP}
\bar{f}(\phi)(\Psi-\Phi)=a^{2}\bar{p}\Pi+\bar{f}_{\phi}\delta\phi+\bar{f}_{\phi}\bar{\phi}'(B-E'),
\end{equation}
where has been used the so-called Bardeen potentials, $\Phi$ and $\Psi$ ~\cite{Bardeen}, which are defined by
\begin{eqnarray}
\Phi&\equiv A+(B-E')'+\mathcal{H}(B-E')\\
\Psi&\equiv\psi-\mathcal{H}(B-E').
\end{eqnarray}
Now, if there is no anisotropic pressure, i.e., if $\Pi=0$, the two potentials can be related to each other as
\begin{equation}\label{eq:relgenpotsinpst}
\Psi=\Phi+\frac{\bar{f}_{\phi}}{\bar{f}}\delta\phi+\frac{\bar{f}_{\phi}}{\bar{f}}\bar{\phi}'(B-E').
\end{equation}
For $\bar{f}=1$, it implies that $\Phi=\Psi$, which corresponds to the case of GR in the absence of anisotropic pressure. We can see, that if we work in the Newtonian gauge, i.e, ($E=B=0$), we obtain of the equation (\ref{eq:relpotsgeneralconP}), the following
\begin{equation}
\bar{f}(\Psi-\Phi)=a^{2}\bar{p}\Pi+\bar{f}_{\phi}\delta\phi,
\end{equation}
or in absence of anisotropic presure
\begin{equation}\label{eq:relpotsinpst1}
\Psi=\Phi+\frac{\bar{f}_{\phi}}{\bar{f}}\delta\phi.
\end{equation}
The perturbed equation of the evolution of the scalar field (\ref{eq:campoST2}) is (see Appendix \ref{sec::xPand})
\begin{eqnarray}\label{eq:pertfriedmannstce}
\fl\bar{\omega}[-\delta\phi''+2\bar{\phi}''A+\nabla^{2}\delta\phi
-2\mathcal{H}\delta\phi'+4\mathcal{H}\bar{\phi}'A+(A'+\nabla^{2}B+3\psi'-\nabla^{2}E')\bar{\phi}']\nonumber\\
\fl-\bar{\omega}_{\phi}(\bar{\phi}''+2\mathcal{H}\bar{\phi}')\delta\phi+\frac{1}{2}a^{2}\bar{f}_{\phi}\delta R+3(\mathcal{H}'+\mathcal{H}^{2})\bar{f}_{\phi\phi}\delta\phi+\frac{1}{2}\bar{\omega}_{\phi}(-2\bar{\phi}'\delta\phi'+2\bar{\phi}^{2}A)-\frac{1}{2}\bar{\omega}_{\phi\phi}\bar{\phi}'^{2}\delta\phi'\nonumber\\
\fl-a^{2}\bar{V}_{\phi\phi}\delta\phi=0,
\end{eqnarray}
where $\delta R$ is (for more details, e.g., see \cite{Hannu_book})
\begin{eqnarray}\label{eq:deltaR2}
&\fl\delta R=a^{-2}[-6\Psi''+2\nabla^{2}(2\psi-A)-6\mathcal{H}(A'+3\psi')-12(\mathcal{H}'+\mathcal{H}^{2})A-2\nabla^{2}B'-6\mathcal{H}\nabla^{2}B\nonumber\\
&+2\nabla^{2}E''+6\mathcal{H}\nabla^{2}E'].
\end{eqnarray}
Now, we find the perturbed Friedmann equation for $f(R)$ theories, starting from the equations (\ref{eq:pertST00})-(\ref{eq:pertSTee}), which are the perturbed Friedmann equations for ST theories with the parameter $\omega(\phi)=0$, under the equivalence relations between both theories.\\
From the equation (\ref{eq:v(phi)enR}), we take
\begin{equation}
\bar{V}_{\phi}=\bar{\phi}\bar{\psi}_{\phi}.
\end{equation}
To relate the above equation in terms of $R$ and $f_{R}$, the equivalence relations (\ref{eq:equivphi}) and (\ref{eq:f(rr)}) are taken, wherewith we get 
\begin{equation}
\bar{V}_{\phi}=\frac{\bar{R}}{2}\bar{f}_{RR},
\end{equation}
where the potential has been rescaled by $\frac{1}{2}$.\\
Thus, the perturbed Friedmann equations of $f(R)$ theories take the form
\begin{eqnarray}\label{eq:pertf(R)00}
&\fl\big[-2\nabla^{2}\psi+6\mathcal{H}\psi'+6\mathcal{H}^{2}A+2\mathcal{H}\nabla^{2}B-2\mathcal{H}\nabla^{2}E'\big]\bar{f}_{R}-3\mathcal{H}^{2}\bar{f}_{RR}\delta R=-a^{2}\delta\rho-\frac{1}{2}a^{2}\bar{R}\bar{f}_{RR}\delta R\nonumber\\
&\fl+3\mathcal{H}(\bar{f}_{RR}\delta R'+\bar{f}^{(3)}_{R}\bar{R}'\delta R)-6\mathcal{H}\bar{f}_{RR}\bar{R}'A-\bar{f}_{RR}\partial^{2}\delta R-\bar{f}_{RR}\bar{R}'(\partial^{2}B+3\psi'-\partial^{2}E'),
\end{eqnarray}
which is the $0-0$ perturbed component,
\begin{eqnarray}\label{eq:pertf(R)te}
&\fl-2(\psi'+\mathcal{H}A)_{,\mu}\bar{f}_{R}=-a^{2}(\bar{\rho}+\bar{p})(v_{,\mu}-B_{,\mu})-\bar{f}_{RR}\partial_{\mu}\delta R'-\bar{f}^{(3)}_{R}\bar{R}'\partial_{\mu}\delta R\nonumber\\
&\fl+A_{,\mu}\bar{f}_{RR}\bar{R}'+\mathcal{H}\bar{f}_{RR}\partial_{\mu}\delta R,
\end{eqnarray}
which is the $0-\mu$ perurbed component, 
\begin{eqnarray}\label{eq:pertf(R)et}
&\fl2(\psi'+\mathcal{H}A-\mathcal{H}'B+\mathcal{H}^{2}B)_{,\mu}\bar{f}_{R}=a^{2}(\bar{\rho}+\bar{p})v_{,\mu}+B_{,\mu}(\bar{f}_{RR}\bar{R}')'-2\mathcal{H}B_{,\mu}\bar{f}_{RR}\bar{R}'
\nonumber\\
&\fl+\bar{f}_{RR}\partial_{\mu}\delta R'+\bar{f}^{(3)}_{R}\bar{R}'\partial_{\mu}\delta R-A_{,\mu}\bar{f}_{RR}\bar{R}'-\mathcal{H}\bar{f}_{RR}\partial_{\mu}\delta R,
\end{eqnarray}
which is the $\mu-0$ perurbed component, and 
\begin{eqnarray}\label{eq:pertf(R)ee}
&\fl-(2\mathcal{H}'+\mathcal{H}^{2})\bar{f}_{RR}\delta R\delta_{\mu\nu}+[2\psi''-\nabla^{2}(\psi-A)+\mathcal{H}(2A'+4\psi')+(4\mathcal{H}'+2\mathcal{H}^{2})A\nonumber\\
&\fl+\nabla^{2}B'+2\mathcal{H}\nabla^{2}B]\bar{f}_{R}\delta_{\mu\nu}-(\nabla^{2}E''+2\mathcal{H}\nabla^{2}E')\bar{f}_{R}\delta_{\mu\nu}\nonumber\\
&\fl+(\psi-A-B'-2\mathcal{H}B+E''+2\mathcal{H}E'),_{\mu\nu}\bar{f}_{R}=\delta pa^{2}\delta_{\mu\nu}+\bar{p}a^{2}\left(\Pi,_{\mu\nu}-\frac{1}{3}\delta_{\mu\nu}\nabla^{2}\Pi\right)\nonumber\\
&\fl-\frac{1}{2}a^{2}\bar{R}\bar{f}_{RR}\delta R\delta_{\mu\nu}+\bar{f}_{RR}\left(\partial_{\mu}\partial_{\nu}-\delta_{\mu\nu}\partial^{2}\right)\delta R+\mathcal{H}\left(\bar{f}_{RR}\delta R'+\bar{f}^{(3)}_{R}\bar{R}'\delta R-2\bar{f}_{RR}\bar{R}'A\right)\delta_{\mu\nu}\nonumber\\
&\fl+\bar{f}_{RR}\bar{R}'\left(B_{,\mu\nu}-2\psi'\delta_{\mu\nu}-E'_{,\mu\nu}\right)+\bar{f}^{(4)}_{R}\bar{R}'^{2}\delta R\delta_{\mu\nu}+2\bar{f}^{(3)}_{R}\bar{R}'\delta R'\delta_{\mu\nu}+\bar{f}^{(3)}_{R}\bar{R}''\delta R\delta_{\mu\nu}\nonumber\\
&\fl+\bar{f}^{(3)}_{R}\bar{R}''\delta R\delta_{\mu\nu}+\bar{f}_{RR}\delta R''\delta_{\mu\nu}-2(\bar{f}^{(3)}_{R}\bar{R}'^{2}+\bar{f}_{RR}\bar{R}'')A\delta_{\mu\nu}+\bar{f}_{RR}\bar{R}'\left(\partial^{2}E'-\partial^{2}B-A'\right)\delta_{\mu\nu},
\end{eqnarray}
is $\mu-\nu$ perturbed component.\\
The off-diagonal part, after having calculated the trace of the above equation
\begin{equation}
\bar{f}_{R}(\Psi-\Phi)=a^{2}\bar{p}\Pi+\bar{f}_{RR}\delta R+\bar{f}_{RR}\bar{R}'(B-E').
\end{equation}
For $\Pi=0$, we take the relations between potentials
\begin{equation}\label{eq:relspotsinpf(R)}
\Psi=\Phi+\frac{\bar{f}_{RR}}{\bar{f}_{R}}\delta R+\frac{\bar{f}_{RR}}{\bar{f}_{R}}\bar{R}'(B-E').
\end{equation}
Taking $f(R)=R$, we get the relations of GR $\Phi=\Psi$ is absence of anisotropic pressure.
%-----------------------------------------------------------------------------
\subsubsection{Vector Perturbations}
Vector metric perturbations  are describes by the line element 
\begin{equation}\label{eq:lineapertvec}
ds^{2}=a^{2}(\eta)\left[-d\eta^{2}-2B^{(V)}_{\mu}d\eta dx^{\mu}+\left(\delta_{\mu\nu}-2E^{(V)}_{(\mu,\nu)}\right)dx^{\mu}dx^{\nu}\right].
\end{equation}
Following the same procedure as the one used to find the scalar perturbations, we obtain the vector perturbations, 
\begin{equation}
\frac{1}{2}\nabla^{2}\left(E'_{\mu}-B_{\mu}\right)\bar{f}(\phi)=a^{2}(\bar{\rho}+\bar{p})(v_{\mu}-B_{\mu}),
\end{equation}
which is the $0-\mu$ perturbed component, 
\begin{eqnarray}
&\fl\left(\frac{1}{2}\nabla^{2}\left(B_{\mu}-E'_{\mu}\right)+2a^{-2}(\mathcal{H}'-\mathcal{H}^{2})B_{\mu}\right)\bar{f}(\phi)=-a^{2}(\bar{\rho}+\bar{p})v_{\mu}-\bar{\omega}(\phi)(\bar{\phi}')^{2}B_{\mu}\nonumber\\
&-\big(\bar{f}_{\phi\phi}(\bar{\phi}')^{2}+\bar{f}_{\phi}\bar{\phi}''\big)B_{\mu}+2\mathcal{H}\bar{f}_{\phi}\bar{\phi}'B_{\mu}.
\end{eqnarray}
which is the $\mu-0$ perturbed component,
\begin{equation}
\fl\left(B'_{(\mu,\nu)}+2\mathcal{H}B_{(\mu,\nu)}-E''_{(\mu,\nu)}-2\mathcal{H}E'_{(\mu,\nu)}\right)\bar{f}(\phi)=-a^{2}\Pi_{(\mu,\nu)}-\bar{f}_{\phi}\bar{\phi}'B_{(\mu,\nu)}+\bar{f}_{\phi}\bar{\phi}'E'_{(\mu,\nu)},
\end{equation}
which is the $\mu-\nu$ perturbed component. The $0-0$ component does not contribute to the vector perturbations.\\
Following the same prodecure for the equivalence between both theories shown in the scalar perturbations, we obtain
\begin{equation}
\frac{1}{2}\nabla^{2}\left(E'_{\mu}-B_{\mu}\right)\bar{f}_{R}=a^{2}(\bar{\rho}+\bar{p})(v_{\mu}-B_{\mu}),
\end{equation}
which is the $0-\mu$ perturbed component,
\begin{eqnarray}
\fl\left(\frac{1}{2}\nabla^{2}\left(B_{\mu}-E'_{\mu}\right)+2a^{-2}(\mathcal{H}'-\mathcal{H}^{2})B_{\mu}\right)\bar{f}_{R}=&-a^{2}(\bar{\rho}+\bar{p})v_{\mu}-\big(\bar{f}^{(3)}_{R}(\bar{R}')^{2}+\bar{f}_{RR}\bar{R}''\big)B_{\mu}\nonumber\\
&+2\mathcal{H}\bar{f}_{RR}\bar{R}'B_{\mu}.
\end{eqnarray}
which is the $\mu-0$ perturbed component, and
\begin{equation}
\fl\left(B'_{(\mu,\nu)}+2\mathcal{H}B_{(\mu,\nu)}-E''_{(\mu,\nu)}-2\mathcal{H}E'_{(\mu,\nu)}\right)\bar{f}_{R}=-a^{2}\Pi_{(\mu,\nu)}-\bar{f}_{RR}\bar{R}'B_{(\mu,\nu)}+\bar{f}_{RR}\bar{R}'E'_{(\mu,\nu)}
\end{equation}
which is the $\mu-\nu$ component.
%-----------------------------------------------------------------------------
\subsubsection{Tensor Perturbations}
The procedure to get to the tensor perturbations is the same as the scalar and vector perturbations. The perturbed equation for the $\mu-\nu$ perturbed component is
\begin{equation}
(E''_{\mu\nu}-\nabla^{2}E_{\mu\nu}+2\mathcal{H}E'_{\mu\nu})\bar{f}(\phi)=a^{2}\Pi_{\mu\nu}-\bar{f}_{\phi}\bar{\phi}'E_{\mu\nu}.
\end{equation}
The other components don't contribute to this perturbation type. Using the equivalences between ST and $f(R)$ theories, we get the perturbed equation for $f(R)$ gravity,
\begin{equation}
(E''_{\mu\nu}-\nabla^{2}E_{\mu\nu}+2\mathcal{H}E'_{\mu\nu})\bar{f}_{R}=a^{2}\Pi_{\mu\nu}-\bar{f}_{RR}\bar{R}'E_{\mu\nu}.
\end{equation}
Perturbed Friedmann equations were found to both theories. It is important write down that the equations to $f(R)$ theories were obtained starting from the equations for ST theories, with the parameter $\omega=0$. Our analysis is general since it is not restricted to a specific gauge. Once the general form for the equivalence was founded, we will concentrate in two examples but now in a specific gauge, in our case we choose the Newtonian-gauge. Henceforth, we are going to use the gauge above mentioned for all our calculations.
%=================================================================================================================================
\subsection{Hu-Sawicki and Starobinsky models}
Below are shown two examples from the equivalences between both $f(R)$ and ST theories under conformal-Newtonian gauge. The first is Hu-Sawicki model ~\cite{Hu-Sawicki}, which is important since it is able of reproducing the accelerated expansion of the universe ~\cite{Kowalski, Simon, Eisenstein} besides to satisfice the tests of the solar system ~\cite{Hu-Sawicki}. Although the reconstruction of the potential has already been studied, we show the equivalences, in the Friedmann equations of the background and the scalar perturbed ones. 
Now, this model is given by ~\cite{Kopp}
\begin{equation}\label{eq:f(R)Hu}
	f(R)=-\frac{2\Lambda}{1+2\frac{\epsilon}{n}(\frac{4\Lambda}{R})^{n}},
\end{equation}
where $\Lambda$ is a constant energy scale whose value coincides with
the measured value $\Lambda=\Lambda_{obs}=3 H^{2}_{0}\Omega_{\Lambda}$ and $\epsilon\ll1$ is a small positive deformation parameter.
We note that the derivative of $f(R)$ is
\begin{equation}\label{eq:f'(R)Hu}
	f_{R}=-\epsilon\left(\frac{4\Lambda}{R}\right)^{n+1}.
\end{equation}
From the equivalence (\ref{eq:equivf(R)}), we have
\begin{equation}\label{eq:RenHu}
	R=4\Lambda\left(\frac{|\phi|}{\epsilon}\right)^{-\frac{1}{n+1}}.
\end{equation}
Thus, rewriting the function (\ref{eq:f(R)Hu}) in terms of the scalar field, gives
\begin{equation}
	f=-2\Lambda+4\Lambda\frac{\epsilon}{n}\left(\frac{|\phi|}{\epsilon}\right)^{\frac{n}{n+1}}.
\end{equation}
Using the equivalence in the potential (\ref{eq:v(phi)enR}). The potential for the Hu-Sawicki model is
\begin{equation}
	V(\phi)=2\Lambda\left(1-2\epsilon\frac{n+1}{n}\left(\frac{|\phi|}{\epsilon}\right)^{\frac{n}{n+1}}\right).
\end{equation}
Once the potential is obtained, we calculate the Friedmann equations in the background in terms of the scalar field, which are 
\begin{equation}
	3\mathcal{H}^{2}\phi=\rho a^{2}-3\mathcal{H}\phi'+2\Lambda\left(1-2\epsilon\frac{n+1}{n}\left(\frac{|\phi|}{\epsilon}\right)^{\frac{n}{n+1}}\right)a^{2}
\end{equation}
and
\begin{equation}
\fl-(2\mathcal{H}'+\mathcal{H}^{2})\phi=p a^{2}+\phi''+\mathcal{H}\phi'-2\Lambda\left(1-2\epsilon\frac{n+1}{n}\left(\frac{|\phi|}{\epsilon}\right)^{\frac{n}{n+1}}\right)a^{2}.
\end{equation}
From the relation (\ref{eq:equivf(R)}), the Friedmann equations in the formalism $f(R)$ take the form
\begin{equation}
\fl-3\mathcal{H}^{2}\epsilon\left(\frac{4\Lambda}{R}\right)^{n+1}=\rho a^{2}+2a^{2}\Lambda\left(1-2\epsilon\frac{n+1}{n}\left(\frac{4\Lambda}{R}\right)^{n}\right)-3\epsilon(n+1)\mathcal{H}\frac{R'}{R}\left(\frac{4\Lambda}{R}\right)^{n+1}
\end{equation}
and
\begin{eqnarray}
\fl(2\mathcal{H}+\mathcal{H}^{2})\epsilon\left(\frac{4\Lambda}{R}\right)^{n+1}=pa^{2}+\epsilon(n+1)\frac{R'}{R}\mathcal{H}\left(\frac{4\Lambda}{R}\right)^{n+1}-2\Lambda\left(1-2\epsilon\frac{n+1}{n}\left(\frac{4\Lambda}{R}\right)^{n}\right)\nonumber\\
\fl+\epsilon(n+1)\frac{R''}{R}\left(\frac{4\Lambda}{R}\right)^{n+1}
-\epsilon(n+1)(n+2)\frac{R'^{2}}{R^{2}}\left(\frac{4\Lambda}{R}\right)^{n+1}.
\end{eqnarray}
The perturbed Friedmann equations (time-time and space-space components) in terms of the scalar field are
\begin{eqnarray}\label{eq:PertSawicki_ST_tt}
\fl[-2\nabla^{2}\Psi+6\mathcal{H}\Psi'+6\mathcal{H}^{2}\Phi]\bar{\phi}-3\mathcal{H}^{2}\delta\phi=-a^{2}\delta\rho+3\mathcal{H}\delta\phi'-	6\mathcal{H}\bar{\phi}'\Phi-\nabla^{2}\delta\phi\nonumber\\
\fl-3\bar{\phi}'\Psi'-4a^{2}\Lambda\left(\frac{|\phi|}{\epsilon}\right)^{-\frac{1}{n+1}}\delta\phi
\end{eqnarray}
and
\begin{eqnarray}\label{eq:PertSawicki_ST_ee}
\fl\big([2\Psi''+\nabla^{2}(\Phi-\Psi)+\mathcal{H}(2\Phi'+4\Psi')+(4\mathcal{H}'+2\mathcal{H}^{2})\Phi]\delta_{\mu\nu}+(\Psi-\Phi),_{\mu\nu}\big)\bar{\phi}\nonumber\\
\fl+(-2\mathcal{H}'-\mathcal{H}^{2})\delta\phi\delta_{\mu\nu}=a^{2}\Big(\delta p \delta_{\mu\nu}+ \bar{p}\left(\Pi,_{\mu\nu}-\frac{1}{3}\delta_{\mu\nu}\nabla^{2}\Pi\right)\Big)+\partial_{\mu}\partial_{\nu}\delta\phi-\nabla^{2}\delta\phi\delta_{\mu\nu}\nonumber\\
\fl+\Big(\delta\phi''-2\bar{\phi}''\Phi-2\mathcal{H}\bar{\phi}'\Phi+\mathcal{H}\delta\phi'-(2\Psi'+\Phi')\bar{\phi}'
-4a^{2}\Lambda\left(\frac{|\phi|}{\epsilon}\right)^{-\frac{1}{n+1}}\delta\phi\Big)\delta_{\mu\nu},
\end{eqnarray}
where we have used
\begin{equation}
V_{\phi}=4\Lambda\left(\frac{|\phi|}{\epsilon}\right)^{-\frac{1}{n+1}}.
\end{equation}
Now, the Friedmann equations in the formalism $f(R)$, using the equivalence relations, take the form
\begin{eqnarray}\label{eq:PertSawicki_f(R)_tt}
\fl-(-2\nabla^{2}\Psi+6\mathcal{H}\Psi'+6\mathcal{H}^{2}\Phi)\epsilon\left(\frac{4\Lambda}{\bar{R}}\right)^{n+1}-3\mathcal{H}^{2}\frac{\epsilon(n+1)}{\bar{R}}\left(\frac{4\Lambda}{\bar{R}}\right)^{n+1}\delta R=-a^{2}\delta\rho\nonumber\\
\fl-\frac{a^{2}\bar{R}}{2}\frac{\epsilon(n+1)}{\bar{R}}\left(\frac{4\Lambda}{\bar{R}}\right)^{n+1}\delta R
-\frac{\epsilon(n+1)}{\bar{R}}\left(\frac{4\Lambda}{\bar{R}}\right)^{n+1}\nabla^{2}\delta R\nonumber\\
\fl+3\mathcal{H}\Big(\frac{\epsilon(n+1)}{\bar{R}}\left(\frac{4\Lambda}{\bar{R}}\right)^{n+1}\delta R'-\epsilon(n+1)(n+2)\frac{\bar{R}'}{\bar{R}^{2}}\left(\frac{4\Lambda}{\bar{R}}\right)^{n+1}\delta R\Big)\nonumber\\
\fl-6\mathcal{H}\frac{\epsilon(n+1)}{\bar{R}}\left(\frac{4\Lambda}{\bar{R}}\right)^{n+1}\bar{R}'\Phi-3\frac{\epsilon(n+1)}{\bar{R}}\left(\frac{4\Lambda}{\bar{R}}\right)^{n+1}\bar{R}'\Psi'.
\end{eqnarray}
and
\begin{eqnarray}\label{eq:PertSawicki_f(R)_ee}
\fl-[[2\Psi''+\nabla^{2}(\Phi-\Psi)+\mathcal{H}(2\Phi'+4\Psi')+(4\mathcal{H}'+2\mathcal{H}^{2})\Phi]\delta_{\mu\nu}+(\Psi-\Phi),_{\mu\nu}]\epsilon\left(\frac{4\Lambda}{\bar{R}}\right)^{n+1}\nonumber\\
\fl-(2\mathcal{H}'+\mathcal{H}^{2})\frac{\epsilon(n+1)}{\bar{R}}\left(\frac{4\Lambda}{\bar{R}}\right)^{n+1}\delta R\delta_{\mu\nu}=a^{2}(\delta p\delta_{\mu\nu}+\bar{p}(\Pi,_{\mu\nu}-\frac{1}{3}\delta_{\mu\nu}\nabla^{2}\Pi))\nonumber\\
\fl-\frac{a^{2}\bar{R}}{2}\frac{\epsilon(n+1)}{\bar{R}}\left(\frac{4\Lambda}{\bar{R}}\right)^{n+1}\delta R\delta_{\mu\nu}+\frac{\epsilon(n+1)}{\bar{R}}\left(\frac{4\Lambda}{\bar{R}}\right)^{n+1}(\partial_{\mu}\partial_{\nu}-\delta_{\mu\nu}\nabla^{2})\delta R\nonumber\\
\fl+\mathcal{H}\Bigg(\frac{\epsilon(n+1)}{\bar{R}}\left(\frac{4\Lambda}{\bar{R}}\right)^{n+1}\delta R'-\bar{R}'\frac{\epsilon(n+1)(n+2)}{\bar{R}^{2}}\left(\frac{4\Lambda}{\bar{R}}\right)^{n+1}\delta R\Bigg)\delta_{\mu\nu}\nonumber\\
\fl-2\mathcal{H}\bar{R}'\frac{\epsilon(n+1)}{\bar{R}}\left(\frac{4\Lambda}{\bar{R}}\right)^{n+1}\Phi\delta_{\mu\nu}+\Bigg(-2\bar{R}'\frac{\epsilon(n+1)(n+2)}{\bar{R}^{2}}\left(\frac{4\Lambda}{\bar{R}}\right)^{n+1}\delta \bar{R}'\nonumber\\
\fl+\bar{R}'^{2}\frac{\epsilon(n+1)(n+2)(n+3)}{\bar{R}^{3}}\left(\frac{4\Lambda}{\bar{R}}\right)^{n+1}\delta R+\frac{\epsilon(n+1)}{\bar{R}}\left(\frac{4\Lambda}{\bar{R}}\right)^{n+1}\delta R''\nonumber\\
\fl-\bar{R}''\frac{\epsilon(n+1)(n+2)}{\bar{R}^{2}}\left(\frac{4\Lambda}{\bar{R}}\right)^{n+1}\delta R\Bigg)\delta_{\mu\nu}-2\Phi\Bigg(-\bar{R}'^{2}\frac{\epsilon(n+1)(n+2)}{\bar{R}^{2}}\left(\frac{4\Lambda}{\bar{R}}\right)^{n+1}\nonumber\\
\fl+\bar{R}''\frac{\epsilon(n+1)}{\bar{R}}\left(\frac{4\Lambda}{\bar{R}}\right)^{n+1}\Bigg)\delta_{\mu\nu}-(\Phi'+2\Psi')\bar{R}'\frac{\epsilon(n+1)}{\bar{R}}\left(\frac{4\Lambda}{\bar{R}}\right)^{n+1}\delta_{\mu\nu},
\end{eqnarray}
where has been used the following relation
\begin{equation}
	\delta\phi=\frac{\epsilon(n+1)}{R}\left(\frac{4\Lambda}{R}\right)^{n+1}\delta R.
\end{equation}
The second model to discuss is the Starobinsky model ~\cite{Starobinsky}, which is a cosmic inflation model. Whose perturbations in the inflationary era were first discussed by Mukhanok and Starobinsky himself ~\cite{Mukhanov2,Starobinsky2}. His predictions agree with the recent CMB data ~\cite{Planck}. For more discussions on this model, see e.g., ~\cite{Artymowski}.\\
The Starobinsky model is given by
\begin{equation}\label{eq:fstaro}
	f(R)=R+\frac{R^{2}}{6M^{2}},
\end{equation}
where the constant $M$ has mass dimenssion. Performing the same above procedure for to construct the potencial, we start from
\begin{equation}\label{eq:f'staro}
	f_{R}=1+\frac{R}{3M^{2}}.
\end{equation}
From the equivalence (\ref{eq:equivf(R)}), we have
\begin{equation}\label{eq:Renstaro}
	R=3M^{2}(\phi-1).
\end{equation}
Thus, the potential gives
\begin{equation}\label{eq:potstaro}
	V(\phi)=\frac{3}{4}M^{2}(\phi-1)^{2},
\end{equation}
where the potential has been rescaled by $\frac{1}{2}$. Found  the potential, we calculate the Friedmann equations in terms of the scalar field
\begin{equation}
	3\mathcal{H}^{2}\phi=\rho a^{2}-3\mathcal{H}\phi'+\frac{3}{4}M^{2}(\phi-1)^{2}a^{2}
\end{equation}
and
\begin{equation}
	-(2\mathcal{H}'+\mathcal{H}^{2})\phi=p a^{2}+\phi''+\mathcal{H}\phi'-\frac{3}{4}M^{2}(\phi-1)^{2}a^{2}.
\end{equation}
From the relation (\ref{eq:equivf(R)}), we obtain the Friedmann equations for the $f(R)$ formalism
\begin{eqnarray}
	3\mathcal{H}^{2}\left(1+\frac{R}{3M^{2}}\right)=\rho a^{2}+a^{2}\frac{R^{2}}{12M^{2}}-\frac{\mathcal{H}R'}{M^{2}}\\
	-(2\mathcal{H}+\mathcal{H}^{2})\left(1+\frac{R}{3M^{2}}\right)=p a^{2}+\frac{\mathcal{H}R'}{3M^{2}}-a^{2}\frac{R^{2}}{12M^{2}}+\frac{R''}{3M^{2}}.
\end{eqnarray}
Now, the perturbed Friedmann equations (time-time and space-space compnents) in terms of the scalar field are
\begin{eqnarray}\label{eq:PertStaro_ST_tt}
\fl[-2\nabla^{2}\Psi+6\mathcal{H}\Psi'+6\mathcal{H}^{2}\Phi]\bar{\phi}-3\mathcal{H}^{2}\delta\phi
=-a^{2}\delta\rho+3\mathcal{H}\delta\phi'-6\mathcal{H}\bar{\phi}'\Phi-\nabla^{2}\delta\phi\nonumber\\
\fl-3\bar{\phi}'\Psi'-\frac{3}{2}M^{2}(\phi-1)a^{2}\delta\phi.
\end{eqnarray}
and
\begin{eqnarray}\label{eq:PertStaro_ST_ee}
\fl\big([2\Psi''+\nabla^{2}(\Phi-\Psi)+\mathcal{H}(2\Phi'+4\Psi')+(4\mathcal{H}'+2\mathcal{H}^{2})\Phi]\delta_{\mu\nu}+(\Psi-\Phi),_{\mu\nu}\big)\bar{\phi}\nonumber\\
\fl+(-2\mathcal{H}'-\mathcal{H}^{2})\delta\phi\delta_{\mu\nu}=a^{2}\Big(\delta p \delta_{\mu\nu}+ \bar{p}\left(\Pi,_{\mu\nu}-\frac{1}{3}\delta_{\mu\nu}\nabla^{2}\Pi\right)\Big)+\partial_{\mu}\partial_{\nu}\delta\phi-\nabla^{2}\delta\phi\delta_{\mu\nu}\nonumber\\
\fl+\Big(\delta\phi''-2\bar{\phi}''\Phi-2\mathcal{H}\bar{\phi}'\Phi
+\mathcal{H}\delta\phi'-(2\Psi'+\Phi')\bar{\phi}'-\frac{3}{2}M^{2}(\phi-1)a^{2}\delta\phi\Big)\delta_{\mu\nu}.
\end{eqnarray}
Once obtained the above perturbations, we find the cosmological perturbations in the formalism $f(R)$, which are
\begin{eqnarray}\label{eq:PertStaro_f(R)_tt}
\fl(-2\nabla^{2}\Psi+6\mathcal{H}\Psi'+6\mathcal{H}^{2}\Phi)\left(1+\frac{\bar{R}}{3M^{2}}\right)-\left(\frac{\mathcal{H}^{2}}{M^{2}}\right)\delta R=-a^{2}\delta\rho-a^{2}\left(\frac{\bar{R}}{6M^{2}}\right)\delta R\nonumber\\
\fl-\left(\frac{1}{3M^{2}}\right)\nabla^{2}\delta R+\left(\frac{\mathcal{H}}{M^{2}}\right)\delta R'
-2\mathcal{H}\left(\frac{\bar{R}'}{M^{2}}\right)\Phi-\left(\frac{\bar{R}'}{M^{2}}\right)\Psi'.
\end{eqnarray}
and
\begin{eqnarray}\label{eq:PertStaro_f(R)_ee}
\fl[[2\Psi''+\nabla^{2}(\Phi-\Psi)+\mathcal{H}(2\Phi'+4\Psi')+(4\mathcal{H}'+2\mathcal{H}^{2})\Phi]\delta_{\mu\nu}+(\Psi-\Phi),_{\mu\nu}]\left(1+\frac{\bar{R}}{3M^{2}}\right)\nonumber\\
\fl-\left(\frac{2\mathcal{H}'+\mathcal{H}^{2}}{3M^{2}}\right)\delta R\delta_{\mu\nu}=a^{2}\delta p\delta_{\mu\nu}+a^{2}\bar{p}(\Pi,_{\mu\nu}-\frac{1}{3}\delta_{\mu\nu}\nabla^{2}\Pi)-a^{2}\left(\frac{\bar{R}}{6M^{2}}\right)\delta R\delta_{\mu\nu}\nonumber\\
\fl+\left(\frac{1}{3M^{2}}\right)(\partial_{\mu}\partial_{\nu}-\delta_{\mu\nu}\nabla^{2})\delta R+\left(\frac{\mathcal{H}}{3M^{2}}\right)\delta R'\delta_{\mu\nu}-\left(\frac{2\mathcal{H}\bar{R}'}{3M^{2}}\right)\Phi\delta_{\mu\nu}+\left(\frac{1}{3M^{2}}\right)\delta R''\delta_{\mu\nu}\nonumber\\
\fl-\left(\frac{2\bar{R}''}{3M^{2}}\right)\Phi\delta_{\mu\nu}-(\Phi'+2\Psi')\left(\frac{\bar{R}'}{3M^{2}}\right)\delta_{\mu\nu}.
\end{eqnarray}
%We show how to obtain the perturbed Friedmann equations (\ref{eq:PertSawicki_f(R)_tt}) and (\ref{eq:PertSawicki_f(R)_ee}) in Hu-Sawicki model for $f(R)$ theories from the equations (\ref{eq:PertSawicki_ST_tt}) and (\ref{eq:PertSawicki_ST_ee}) for ST theories and for the Starobinsky model we get (\ref{eq:PertStaro_f(R)_tt}) and (\ref{eq:PertStaro_f(R)_ee}) for $f(R)$ theories from the equations (\ref{eq:PertStaro_ST_tt}) and (\ref{eq:PertStaro_ST_ee}) for ST theories
We show how to obtain the perturbed Friedmann equations for Hu-Sawicki and Starobinsky models for $f(R)$ theories from the perturbed equations for ST theories. In the next subsection, the linear evolution of matter density perturbations under the sub-horizon approximation in the conformal-Newtonian gauge for each of the theories will be calculated.
%=================================================================================================================================
\subsection{Sub-horizon Approximation in ST and $f(R)$  theories under conformal-Newtonian gauge}\label{sec::sub-horizon}
The perturbed energy conservation equations are ~\cite{Hannu_book}
\begin{equation}\label{eq:conenergiapert1}
	\delta'=(1+w)(\nabla^{2}v+3\Psi')+3\mathcal{H}(w\delta-\frac{\delta p}{\bar{\rho}})
\end{equation}
and
\begin{equation}\label{eq:conmomentumpert1}
	v'=-\mathcal{H}(1-3w)v-\frac{w'}{1+w}v+\frac{\delta p}{\bar{\rho}(1+w)}+\frac{2}{3}\frac{w}{1+w}\nabla^{2}\Pi+\Phi,
\end{equation}
where $\delta=\frac{\delta\rho}{\bar{\rho}}$ is the perturbation of the relative energy density  and $v$ is the perturbation of velocity.\\
The above equations in the matter domain, i.e., $w=0$ (do not confuse with the parameter $\omega$ of ST theories) and taking $\Pi=0$, takes the form in the space Fourier as
\begin{equation}
	\delta_{m}''+\mathcal{H}\delta_{m}'+k^{2}\Phi-3\mathcal{H}\Psi'-3\Psi''=0,
\end{equation}
where $k$ is the wave number. Taking the sub-horizon approach, i.e., $\frac{\partial}{\partial\eta}\sim\mathcal{H}\ll k$. we get
\begin{equation}\label{eq:materiasubhubble}
	\delta_{m}''+\mathcal{H}\delta_{m}'+k^{2}\Phi=0.
\end{equation}
To obtain the Poisson type equation in the sub-horizon approach for ST theories, the perturbed time-time component (\ref{eq:pertST00}) in the Fourier space is taken  
\begin{equation}\label{eq:friedmannst_t-t_sub}
	2k^{2}\Psi\bar{f}=-a^{2}\delta\rho+k^{2}\bar{f}_{\phi}\delta\phi.
\end{equation}
To find $\delta\phi$ of the above equation, we take $\bar{\omega}$ constant in the equation (\ref{eq:pertfriedmannstce}), thus
\begin{equation}\label{eq:deltaphisubhorizonte}
	-\bar{\omega}k^{2}\delta\phi+\frac{1}{2}a^{2}\bar{f}_{\phi}\delta R=0.
\end{equation}
Applying the sub-horizon approach to the term $\delta R$ (\ref{eq:deltaR2}), we get
\begin{equation}\label{eq:deltaRsub}
	\delta R=-2a^{-2}k^{2}(2\Psi-\Phi),
\end{equation}
replacing $\delta R$ in the equation (\ref{eq:deltaphisubhorizonte}), gives
\begin{equation}
	\delta\phi=\frac{\bar{f}_{\phi}}{\bar{\omega}}(\Phi-2\Psi).
\end{equation}
Using the relation (\ref{eq:relpotsinpst1}) in the above equation
\begin{equation}\label{eq:deltaphiensub}
	\delta\phi=-\frac{\bar{f}_{\phi}}{\left(\bar{\omega}+2\frac{\bar{f}_{\phi}^{2}}{\bar{f}}\right)}\Phi.
\end{equation}
It can be seen that the perturbations of the scalar field in ST gravity theories do not depend on the wave number $k$ in the sub-horizon approach. Replacing the above expression and using (\ref{eq:relpotsinpst1}) in (\ref{eq:friedmannst_t-t_sub}), we get Poisson type equation
\begin{equation}
	k^{2}\Phi=-4\pi G^{ST}_{eff}a^{2}\bar{\rho}_{m}\delta_{m},
\end{equation}
where
\begin{equation}
	G^{ST}_{eff}=\frac{1}{8\pi\bar{f}}\left(\frac{2\bar{\omega}+4\frac{\bar{f}_{\phi}^{2}}{\bar{f}}}{2\bar{\omega}+3\frac{\bar{f}_{\phi}^{2}}{\bar{f}}}\right),
\end{equation}
is the gravitational effective constant for ST theories.\\
The linear evolution of matter density perturbations and scalar field perturbations in sub-horizon approach in the framework of ST
theory of gravity can be written as follows
\begin{equation}
	\delta_{m}''+\mathcal{H}\delta_{m}'-4\pi G^{ST}_{eff}a^{2}\bar{\rho}_{m}\delta_{m}=0,
\end{equation}
where it has been used (\ref{eq:materiasubhubble}). For more details about the density linear perturbations see ~\cite{Boisseau,Nazari-Pooya,Sanchez}.\\
To calculate the perturbations in $f(R)$ theory under the same approach, we start of the perturbed time-time component (\ref{eq:pertf(R)00})
\begin{equation}
	2k^{2}\Psi\bar{f}_{R}=-a\delta\rho_{m}+k^{2}\bar{f}_{RR}\delta R,
\end{equation}
replacing the relation (\ref{eq:relspotsinpf(R)}) in the above equation, we take
\begin{equation}
	k^{2}\Phi=-\frac{a^{2}}{2\bar{f}_{R}}\delta\rho_{m}-\frac{k^{2}}{2}\frac{\bar{f}_{RR}}{\bar{f}_{R}}\delta R.
\end{equation}
Using the equation (\ref{eq:deltaRsub}), we get
\begin{equation}
	k^{2}\Phi=-4\pi G^{f(R)}_{eff}a^{2}\bar{\rho}_{m}\delta_{m}.
\end{equation}
This is the Poisson equation in the Fourier space in $f(R)$ theories, where
\begin{equation}\label{eq:Genf(R)}
	G^{f(R)}_{eff}=\frac{1}{8\pi \bar{f}_{R}}\left(\frac{1+4\frac{k^{2}}{a^{2}}\frac{\bar{f}_{RR}}{\bar{f}_{R}}}{1+3\frac{k^{2}}{a^{2}}\frac{\bar{f}_{RR}}{\bar{f}_{R}}}\right),
\end{equation}
is the gravitational effective constant for $f(R)$ theories. The linear evolution of matter density perturbations in sub-horizon approach for $f(R)$ theories of gravity is
\begin{equation}
	\delta_{m}''+\mathcal{H}\delta_{m}'-4\pi G^{f(R)}_{eff}a^{2}\bar{\rho}\delta_{m}=0,
\end{equation}
where has ben used (\ref{eq:materiasubhubble}). For more detail about the linear density perturbations in $f(R)$ theories see ~\cite{delaCruz,Chiu,Tsujikawa}.\\
To see the equivalences in both theories, we take $\bar{\omega}=0$ in the effective gravitational constant for ST theories and let's use the equation (\ref{eq:equivf(R)}) to obtain
\begin{equation}
	G_{eff}=\frac{4}{3}\frac{1}{8\pi\bar{f}_{R}},
\end{equation}
if we use $\frac{k^{2}}{a^{2}}\frac{\bar{f}_{RR}}{\bar{f}_{R}}\gg1$ in (\ref{eq:Genf(R)}),
we obtain the same effective gravitational constant ~\cite{Tsujikawa}. In this way, we show the equivalence between $f(R)$ and ST theories under the sub-horizon approximation using the above limit. Finally, if we take the variation of $\delta R$ (\ref{eq:deltaRsub}) and we use the relation (\ref{eq:relspotsinpf(R)}), we obtain
\begin{equation}
	\delta R=-\frac{2\frac{k^{2}}{a^{2}}}{1+4\frac{k^{2}}{a^{2}}\frac{\bar{f}_{RR}}{\bar{f}_{R}}}.
\end{equation}
Now, using the above limit, the expression yields
\begin{equation}
	\delta R=-\frac{\bar{f}_{R}}{2\bar{f}_{RR}}\Phi.
\end{equation}
If we apply the equivalence in (\ref{eq:deltaphiensub}), i.e., taking $\bar{\omega}=0$ and using (\ref{eq:equivphi}), (\ref{eq:equivf(R)}) we obtain the same above expression. Evidencing once again the  equivalence between both theories
%===============================================================================================================================================
\section{\label{sec::conclusiones}Summary and conclusions}
In this section we present the results found in this paper
\begin{itemize}
	\item Using the variational principle to obtain the field equations in the metric formalism we have used  the Gibbons-York-Hawking boundary term type to make no further assumptions about variations of the metric $\delta g_ {ab}$, at the boundary. Furthermore, to obtain the field equation for the scalar field, it is necesary to impose, $\delta\phi=0$ at the boundary. Following ~\cite{Alejo}, where they found the field equations for $f(R)$ and they showed that, in addition to $\delta g_{ab}$, it is necesary to impose that $\delta R=0$ at the boundary. It allow us to reinforced equivalences between both theories.
	\item  We found the scalar, vector and tensor cosmological perturbations for ST theories. Then, using the equivalences between both theories we obtained the perturbed ones, for f(R) gravity in any gauge.
	\item We showed how to obtain the potential for the Hu-Sawicki and Starobinsky $f(R)$ models. Then, we calculated the Friedmann equations for the background and perturbed universe in terms of the scalar field for these models under the conformal-Newtonian gauge, and 	taking into account the equivalence between theories, we find the Friedmann equations for the $f(R)$ formalism.
	\item The Poisson type equations and linear evolution of matter density perturbations for both ST (with the parameter $\omega$ constant) and $f(R)$ theories were obtained under the conformal-Newtonian gauge. We showed the equivalences between effective gravitational constant for both theories, using for ST the parameter $\omega=0$ and using the limit $\frac{k^{2}}{a^{2}}\frac{\bar{f}_{RR}}{\bar{f}_{R}}\gg1$ for $f(R)$ theories.
	%\item \textcolor{blue}{Lo interesante de las equivalencias, es que dependiendo del modelo, se puede analizar mejor las ecuaciones en términos del campo escalar que con las ecuaciones de las teorías $f(R)$}.
	\item We showed how to obtain the perturbations for the ST theores in the package \textit{xPAnd} under the software \textbf{Mathematica}.
\end{itemize}
%============================================================================================================================================================================================================================================
\appendix
\section{\label{AppendixA} Terms with $M_{c}$ and $N^{c}$}
As a alreeady mentioned, the quantities $M_{c}$ and $N^{c}$ are defined by
\begin{equation}
	M_{c}=\frac{f(\phi)}{2} g_{ef}\nabla_{c}(\delta g^{ef})-\frac{1}{2}(\delta g^{ef})g_{ef}\nabla_{c}f(\phi)
\end{equation}
and
\begin{equation}
	N^{c}=\frac{f(\phi)}{2}\nabla_{f}(\delta g^{cf})-\frac{1}{2}(\delta g^{cf})\nabla_{f}f(\phi).
\end{equation}
The covariant derivative of $M_{c}$ is
\begin{eqnarray*}
	\nabla^{c}M_{c}&=\frac{1}{2}\nabla^{c}(f(\phi) g_{ef}\nabla_{c}\delta
	g^{ef})-\frac{1}{2}\nabla^{c}(\delta g^{ef}g_{ef}\nabla_{c}f(\phi))\nonumber\\
	&=\frac{1}{2}(\nabla^{c}f(\phi))g_{ef}\nabla_{c}\delta
	g^{ef}+\frac{1}{2}f(\phi) g_{ef}\nabla^{c}\nabla_{c}\delta
	g^{ef}\nonumber\\
	&\quad-\frac{1}{2}(\nabla^{c}\delta g^{ef})g_{ef}\nabla_{c}f(\phi)-\frac{1}{2}\delta
	g^{ef}g_{ef}\nabla^{c}\nabla_{c}f(\phi),
\end{eqnarray*}
where the first and third term are canceled, with which
\begin{equation}
	\label{eq:partes1}
	\nabla^{c}M_{c}=\frac{1}{2}f(\phi) g_{ef}\Box\delta
	g^{ef}-\frac{1}{2}\delta g^{ef}g_{ef}\Box f(\phi),
\end{equation}
so
\begin{equation}
	\label{eq:partes2}
	\frac{1}{2}f(\phi) g_{ef}\Box\delta
	g^{ef}=\nabla^{c}M_{c}+\frac{1}{2}\delta g^{ef}g_{ef}\Box f(\phi).
\end{equation}
The covariant derivative of $N^{c}$ is
\begin{eqnarray*}
	\nabla_{c}N^{c}&=\frac{1}{2}\nabla_{c}(f(\phi)\nabla_{f}\delta g^{cf})-\frac{1}{2}\nabla_{c}(\delta
	g^{cf}\nabla_{f}f(\phi))\\
	&=\frac{1}{2}(\nabla_{c}f(\phi))\nabla_{f}\delta g^{cf}+\frac{1}{2}f(\phi)\nabla_{c}\nabla_{f}\delta
	g^{cf}\\
	&\quad-\frac{1}{2}(\nabla_{c}\delta g^{cf})\nabla_{f}f(\phi)-\frac{1}{2}\delta
	g^{cf}\nabla_{c}\nabla_{f}f(\phi),
\end{eqnarray*}
where the first and third term are canceled, thus
\begin{equation}
	\label{eq:partes3}
	\nabla_{c}N^{c}=\frac{1}{2}f(\phi)\nabla_{c}\nabla_{f}\delta
	g^{cf}-\frac{1}{2}\delta g^{cf}\nabla_{c}\nabla_{f}f(\phi)
\end{equation}
getting
\begin{equation}
	\label{eq:partes4}
	\frac{1}{2}f(\phi)\nabla_{c}\nabla_{f}\delta
	g^{cf}=\nabla_{c}N^{c}+\frac{1}{2}\delta g^{cf}\nabla_{c}\nabla_{f}\phi.
\end{equation}
Subtracting (\ref{eq:partes2}) with (\ref{eq:partes4}), we get
\begin{eqnarray}
	&\frac{f(\phi)}{2}(g_{ef}\Box\delta g^{ef}-\nabla_{c}\nabla_{f}\delta
	g^{cf})=\nonumber\\ 
	&\frac{1}{2}\delta g^{ef}(g_{ef}\Box f(\phi)-\nabla_{e}\nabla_{f}f(\phi))+(\nabla^{c}M_{c}-\nabla_{c}N^{c}),
\end{eqnarray}
%=================================================================================================================================================
\section{\label{sec::xPand}Cosmological Perturbations in xPAnd}
To obtain the cosmological perturbations, the package \textit{xPand} was used (for more details of the package see ~\cite{Pitrou}). \\
We upload the notebook on  https://github.com/JoelVelasquez/Cosmological-perturbations, where we show how to calculate the cosmological perturbations in ST and f(R) gravity.

\section*{References}
\bibliography{references}

\end{document}